\documentclass[iop]{emulateapj}
\usepackage{verbatim}
\usepackage{rotating}
\usepackage{graphicx}
\usepackage{natbib}

\begin{document}
\title{A Systematic Analysis of Caustic Methods for Galaxy Cluster Masses}
\author{Daniel Gifford, Christopher Miller and Nicholas Kern}
\affil{Department of Astronomy, University of Michigan, 500 Church St. Ann Arbor, MI USA}

\begin{abstract}

We quantify the expected observed statistical and systematic uncertainties of the escape velocity as a measure of the gravitational potential and total mass of galaxy clusters. We focus our attention on low redshift (z $\le 0.15$) clusters, where large and deep spectroscopic datasets currently exist.  Utilizing a suite of Millennium Simulation semi-analytic galaxy catalogs, we find that the dynamical mass, as traced by either the virial relation or the escape velocity, is robust to variations in how dynamical friction is applied to ``orphan'' galaxies in the mock catalogs (i.e., those galaxies whose dark matter halos have fallen below the resolution limit). We find that the caustic technique recovers the known halo masses ($M_{200}$) with a third less scatter compared to the virial masses. The bias we measure increases quickly as the number of galaxies used decreases. For N$_{gal} > 25$, the scatter in the escape velocity mass is dominated by projections along the line-of-sight. Algorithmic uncertainties from the determination of the projected escape velocity profile are negligible. We quantify how target selection based on magnitude, color, and projected radial separation can induce small additional biases into the escape velocity masses. Using $N_{gal} = 150 (25)$, the caustic technique has a per cluster scatter in $\ln(M|M_{200})$ of $0.3$ $(0.5)$ and bias $1\pm{3\%} (16\pm{5\%})$  for clusters with masses $> 10^{14}M_{\odot}$ at $z<0.15$.

\end{abstract}

\section{Introduction}
\label{sec:intro}

Upcoming galaxy cluster surveys from the millimeter to the X-ray wavelengths have the potential to identify thousands to hundreds of thousands of groups and clusters \citep{Song12,Pillepich12}. These large cluster samples will be used to constrain the cosmological parameters which govern the growth of structure in our Universe \citep{Miller01a, Miller01b, Rozo09a, Vikhlinin09, Benson11}. Regardless of the luminous tracer used to infer the underlying dark matter distribution (e.g., normal galaxies, emission-line galaxies,  luminous red galaxies, quasi-stellar objects, galaxy clusters), our ability to constrain the cosmological parameters depends on the accuracy and precision of the mass.

Unlike most galaxies, the masses of galaxy clusters can be directly inferred via the observational signature of the gravitational potential (or its derivative). In fact, clusters are the only object for which current technology enable three physically independent mass estimation techniques: via the dynamics of the member galaxies, via the hot gas in the intra-cluster medium, and via gravitational lensing. These three techniques provide a vital cross-check on the mass estimation techniques, assuming one can quantify the statistical precision and accuracy of the cluster mass estimates themselves. 

Cosmological N-body and hydrodynamic simulations play a new and important role in characterizing the statistical and systematic uncertainties on cluster mass estimates \citep{Nagai07, Becker11}. There has been excellent recent progress on the important step of utilizing realistic mock astronomical observations based on ideal simulations \citep{Meneghetti12, Rasia12, Saro12}. Our primary goal is to utilize a diverse suite of semi-analytic galaxy catalogs to study how well a realistic spectroscopic program can constrain the dynamical masses of galaxy clusters in the low redshift Universe. We quantify our results based on the scatter and bias of the ``observed'' dynamical mass when compared to the halo masses M$_{200}$, which refers to the mass within a radius of $r_{200}$.

When inferring the dynamical masses of clusters, we require an accurate measurement of the galaxy peculiar velocities. With their total dispersion, their dispersion profiles, and/or their escape velocity profile, we can infer cluster masses based on the virial theorem, the Jeans relation for a collision-less fluid, or the caustic technique. To first order, galaxies dynamically respond to the influence of the Newtonian gravitational potential, regardless of their luminosities, shapes, colors, or star-formation histories. Yet galaxies traveling within a cluster environment are likely to have had one or more localized and short-lived gravitational interaction over its lifetime. This ``dynamical friction'' alters the total cluster velocity distribution away from its simple Newtonian expectations, an effect that needs to be captured by the semi-analytic galaxies. In this work, we use sub-halo catalogs as well as a suite of semi-analytic mock galaxy catalogs, to explore how sensitive dynamical masses are to different prescriptions of this dynamical friction.

Given the above context and the state of both cosmological simulations and semi-analytic galaxy formation, the question as to whether these mock galaxies capture the true velocity distribution of real galaxies in clusters is still unanswered. However, the current simulated data now contain a wide variety of options for how galaxies are pasted into the clusters. More-so, these semi-analytic techniques create mock galaxy clusters which very closely resemble the observed Universe in many respects \citep{Guo11}. While the semi-analytic galaxy catalogs are not fully simulated Universes, they do provide us with an opportunity to assess how well our theories and algorithms could do under realistic observational conditions (e.g., non-ideal target selection). 

In this study, we focus on how different tracers in the N-body simulations (e.g. sub-halos or semi-analytic galaxies) allow us to observe the gravitational potential and measure the projected escape velocity to infer the cluster mass. This technique is analogous to applying the Jeans equation, except that the cluster observable is the radial escape velocity as opposed to the velocity dispersion and that the escape velocity maps directly to the gravitational potential, whereas the Jeans analysis maps to its derivative. These differences are subtle but important. Regardless, both the Jean's technique and the caustic technique posit that the radius/velocity {\it phase-space} does indeed map directly to the gravitational potential and through some simplifying assumptions, ultimately to the gravitational mass.

Our primary goal is to present the  statistical characterization (accuracy and precision) of caustic inferred halos masses, as well as study the effects of survey strategy when planning spectroscopic follow-up. In Section 2 we discuss the caustic technique in detail and apply this technique in Section 3 on N-body simulations using the underlying particles, the sub-halos, as well as on the semi-analytic mock galaxy catalogs. Using the galaxy catalogs, we incorporate realistic targeting scenarios and show the effects on the measured bias and scatter.

\section{Methods and Data} \label{Methods}
    \subsection{Inferring Halo Mass from Gravitational Potential}
    \label{sec:methods_caustics}
        Under Newtonian dynamics, the escape velocity is related to the gravitational potential of the system,
        \begin{equation}
            v_{esc}^2(r) = -2\Phi(r) .
        \label{eq:escape}
        \end{equation}
        If the dynamics of the system are controlled by the gravitational potential, tracers which have not escaped the potential well should exist in a well-defined region of $r-v$ phase space, where $r$ is the physical or projected radius from the center of the cluster and $v$ is the peculiar 3-dimensional velocity or projected 1-dimensional velocity respectively relative to the bulk cluster motion. The edge of this region in $r-v$ space within which bound tracers are allowed to exist defines the escape velocity, $v_{esc}(r)$.
        
        In Figure 1, we show an example halo from the Millennium Simulation where we identify the actual gravitational potential of the dark matter $G \Sigma_i \frac{m}{|x -x_i|}$ (red lines) and the iso-density contour which traces the escape velocity profile of the halo (blue lines). In the left panel, the velocities and radii are 3-dimensional and in spherical coordinates while in the right panel they are projected along one line-of-sight. The surface that defines the density edge in the $r-v$ phase space is an iso-density contour that follows $v_{esc}(r)$ and therefore $\Phi(r)$.
        
        \begin{figure*}
        \plottwo{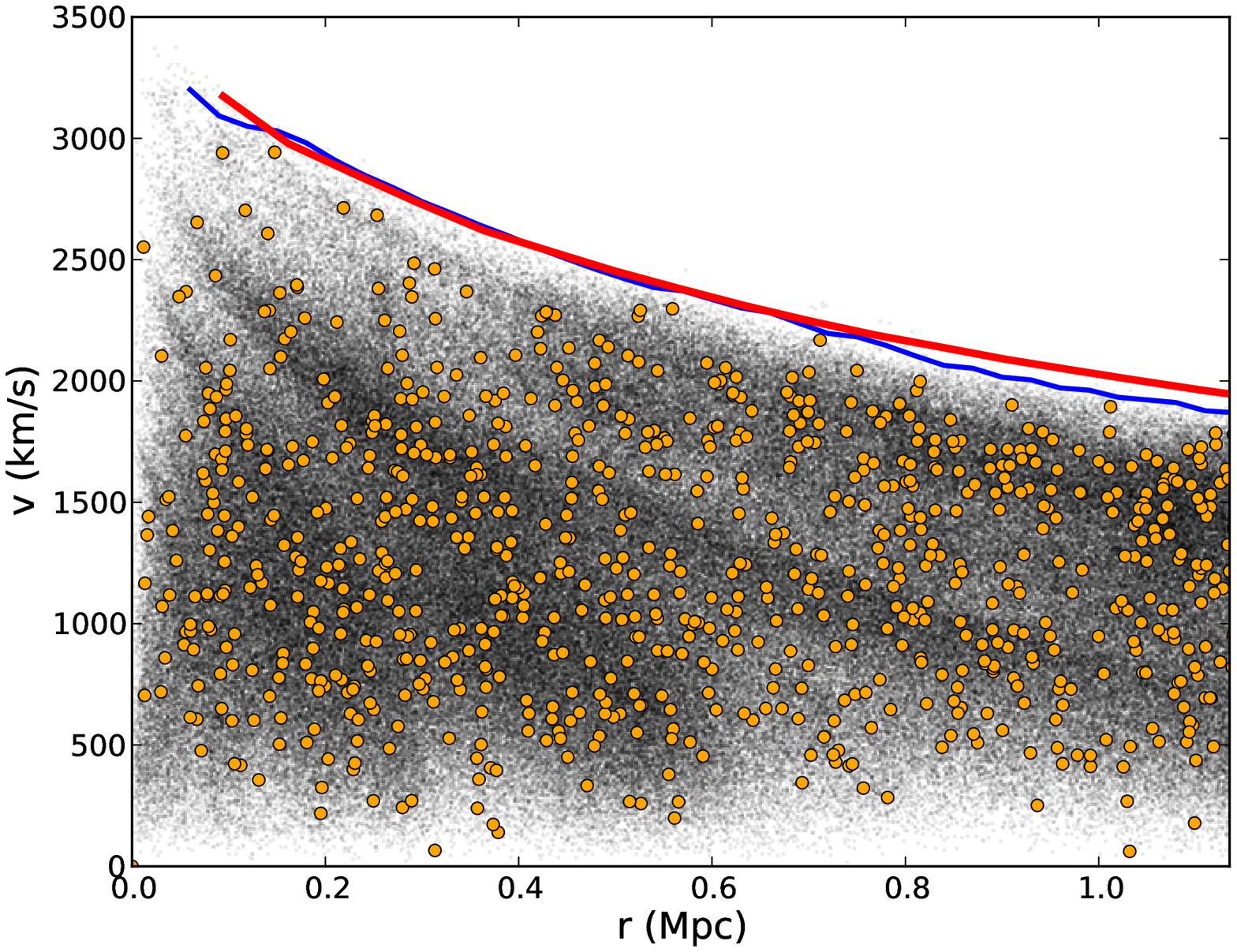}{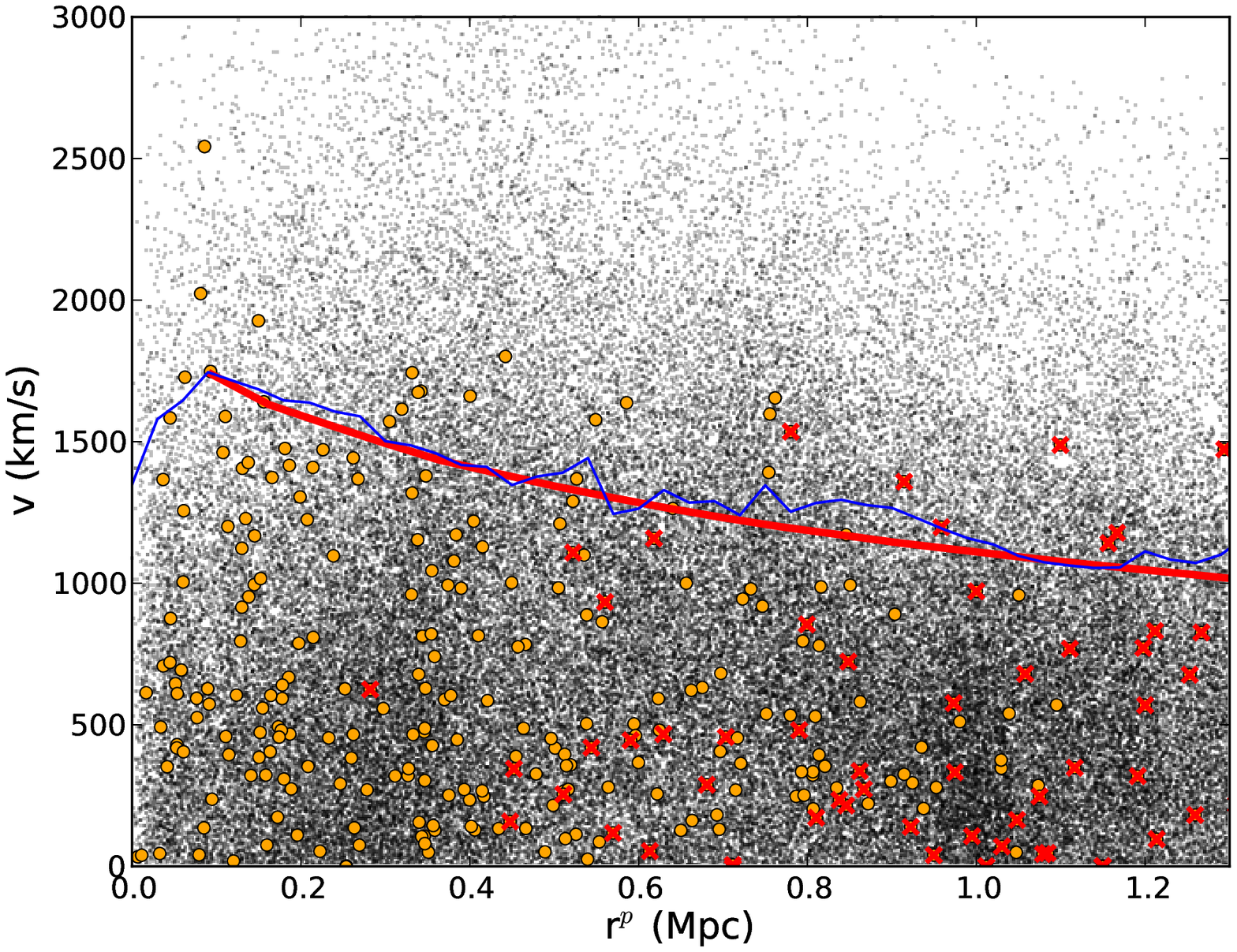}
        \caption{{\bf Left}: The gravitational potential (red band) is shown to envelope the edge of the particle (black points) and galaxy (orange circles) data when projected in the 3D radius-redshift space. The edge of the phase-space density can be defined by choosing the correct iso-density contour (blue). {\bf Right}: The same halo projected on the sky, which blurs the surface from both the positions and anisotropies in the velocity components. Galaxies that are projected into the space, but live outside the virial radius in 3-dimensions are highlighted with red x's.  
        \label{fig:caustics}}
        \end{figure*}

        In observed data, we identify the projected $v_{esc}(r)$ surface by applying standard kernel density estimation techniques to the dynamical tracers in the $r-v$ phase-space. The observed tracers have inherent observational uncertainties in both the radial and velocity directions. In this work, we focus on low-redshift SDSS-like observations with spectroscopic precision of $\sim 50 \rm{km\ s}^{-1}$ or $0.5 h^{-1} \rm{Mpc}$ in normalized coordinates and astrometric precision of  $0.05 h^{-1} \rm{Mpc}$. Therefore, our kernel must be non-symmetric to account for the factor of ten difference in the two dimensions of the phase-space. \citet{Geller99} showed that such axis weighting does not have a large effect on the mass profile determination, something that we confirm in this work. We use a fixed multi-dimensional gaussian kernel with a width in the $r$ and $v$ directions that independently adapt to the sampling according to (Silverman 1998)
        \begin{equation}
            \rm{K(}r,v\rm{)} = \left(\frac{4}{3 N}\right )^{1/5}\sigma_{r,v}
            \label{eq:kernel}
        \end{equation}
        where $N$ is the number of dynamical tracers in the total phase-space and $\sigma_{r,v}$ is the dispersion in the radial and velocity dimensions. Equation \ref{eq:kernel} minimizes the mean integrated squared error of the density estimate, which is the sum of the square of the statistical bias and the variance, also known as the statistical risk \citep{Stien81, Miller02}). While \citet{Diaferio99} adopt an adaptive kernel technique, we will show that a standard fixed kernel recovers the cluster mass estimates with low scatter and bias.
        
        \citet{Diaferio99} assert that any realistic models of galaxy clusters exhibit escape velocity profiles that at no point exceed $\frac{d \ln v_{esc}}{d \ln r} = \zeta$ where $\zeta = 1/4$. If an iso-density contour breaks this limit along its surface, the $v_{esc}(r)$ value is replaced with a new value that yields $\frac{d \ln v_{esc}}{d \ln r} = \zeta$. Here, we follow the prescription used in \citet{Serra11} which invoke a looser constraint of $\zeta = 2$ rather than $\zeta = 1/4$. This allows the algorithm to remove very drastic changes in $v_{esc}$, but does not overly restrict the iso-density contour values.
        
        Once the iso-density contours are determined we must choose which surface corresponds to the escape velocity. However, here we follow the standard procedure of assuming that the data are for time-averaged, self-gravitating, isolated clusters in a steady state. As described in \cite{Binney87}, virialization means that the average system kinetic energy is half of the average system potential energy (KE = $\alpha$PE and $\alpha=0.5$) within the virial radius $r_{vir}$. In combination with Equation \ref{eq:escape} this leads to 
        \begin{equation}
        \langle v_{esc}^2(r<r_{vir}) \rangle - 4 \langle v^2(r<r_{vir}) \rangle = 0
        \label{eq:vesc_sigmav}
        \end{equation}
        Equation \ref{eq:vesc_sigmav} is defined for three dimensional measurements of the velocities. \cite{Gifford13b} test this virialization condition in the Millennium Simulation and find it to hold when the system average energies are calculated near the virial radius.
        
    In real data we only observe the projected line-of-sight component of both the velocity dispersion and the escape velocity. There could exist some level of anisotropy in the velocity vectors: for example the radial and non-radial components of the velocity are not equal: $\langle v_{\theta} \rangle  = \langle v_{\phi} \rangle \ne  \langle v_{r} \rangle$. This is parameterized by the anisotropy parameter $\beta(r) = 1 - \frac{\langle v^2_{\theta}\rangle(r)}{\langle v^2_r\rangle(r)}$. By assuming that the escape velocity profile we measure in projection $v_{esc,los}^2(r_{\perp}) = v_{esc,\theta}(r_{\perp})$, where $r_{\perp}$ is where the 3-dimensional radius $r$ equals the projected radius $r^{p}$, \citet{Diaferio99} show that
        \begin{equation}
        \langle v^2_{esc,los}\rangle (r_{\perp}) = \frac{(1-\beta(r_{\perp}))}{(3-2\beta(r_{\perp}))}\langle v_{esc}^2(r_{\perp}) \rangle \approx   g(\beta(r)) \langle v_{esc}^2(r) \rangle
        \label{eq:v_proj}
        \end{equation}
        and as in  \citet{Diaferio99}, we define the projection correction term:
        \begin{equation}
        g(\beta(r)) = \frac{3-2\beta(r)}{1-\beta(r)}.
        \label{eq:gbeta}
        \end{equation}

        Recalling that equation \ref{eq:vesc_sigmav} is for the three-dimensional velocities, it can be re-written in terms of the observed projected quantities (i.e., the projected radii $r^p$ and the line-of-sight velocities of the galaxies):
        \begin{equation}
        \langle \langle v_{los,esc}^2 \rangle (r^p<r^p_{vir}) \rangle - 4 \langle v^2_{los}(r^p<r^p_{vir}) \rangle = 0
        \label{eq:vesc_sigmav_proj}
        \end{equation}      

        We calculate the velocity dispersion $\sqrt{\langle v_{los}^2 \rangle}$ (hereafter labeled  $\sigma^{v}$) using a robust median-weighted technique on galaxies within the projected radius that has $\frac{\delta \rho}{\rho_{crit}} = 200$ (i.e., $r_{200}$), which is $\langle v^2 \rangle_{r_{200}}$. Interlopers are removed via a shifting-gapper technique described in \S \ref{sec:interloper_removal}.
        
        We then choose the iso-density contour that satisfies equation \ref{eq:vesc_sigmav_proj}. There is some uncertainty in the determination of this surface, which we quantify in \S \ref{sec:results_los}. We note that we always use the same tracers when calculating the dispersions and the phase-space density

        Ideally, we could use the escape velocity profile, which we assume to be the potential profile through equation \ref{eq:escape}, to estimate a mass by using the Poisson equation $\nabla^2 \Phi(x) = 4 \pi G \rho(x)$ or some variation to arrive at the mass profile. However, this ideal scenario involves the challenge of taking derivatives of a noisy estimate of $\Phi (r)$. Instead, \citet{Diaferio97} introduce an alternative estimation through the partial mass differential equation $dm = 4 \pi \rho (r) r^2 dr$. Invoking equation \ref{eq:escape}, we may rewrite this differential as:
        \begin{equation}
        dm = -2 \pi v_{esc}^2(r) \frac{\rho (r) r^2}{\Phi (r)} dr \nonumber
        \end{equation}
        and integrate to arrive at:
        \begin{equation}
        G M(<R) = \int_0^R -2 G \pi v_{esc}^2(r) \frac{\rho (r) r^2}{\Phi (r)} dr
        \label{eq:m_caustic_full}
        \end{equation} 
        
       After identifying the iso-density contour that describes the projected $\langle v_{esc,los}\rangle (r)$, we now have an estimate for $\Phi(r)$ by using equation \ref{eq:escape} and $g(\beta(r))$ and our equation now becomes:
        \begin{eqnarray}
        G M(<R) &=& \int_0^R -2 G \pi g(\beta(r))\langle v_{los,esc}^2\rangle (r) \frac{\rho (r) r^2}{\Phi (r)} dr \\
        G M(<R) &=& \int_0^R \mathcal{F}_{\beta}(r) \langle v_{los,esc}^2 \rangle (r) dr
        \label{eq:caustic_final}
        \end{eqnarray}
        where
        \begin{equation}
        \mathcal{F}_{\beta}(r) = -2 G \pi g(\beta(r)) \frac{\rho (r) r^2}{\Phi (r)} 
        \end{equation}
        
        \citet{Diaferio99} claim that $\mathcal{F}_{\beta}(r)$ is roughly constant as a function of radius from ~1-3$r_{200}$ as calibrated against simulations. For instance, \citet{Diaferio99} find $\langle \mathcal{F}_{\beta}(r) \rangle = 0.5$ and \citet{Serra11} find $\langle \mathcal{F}_{\beta}(r) \rangle = 0.7$.   We use a constant value calibrated against the Millennium Simulations to be $\mathcal{F}_{\beta} = 0.65$ and discuss the implications of both the assumption of a constant as well as its calibration in Section \ref{sec:Discussion}. For a variation of the caustic technique that does not require this calibration, see \cite{Gifford13b}.

    \subsection{Inferring Halo Masses from the Virial Relation}
    \label{sec:methods_virial}
    
\citet{Evrard08} show that the velocity dispersion of a dark matter halo obeys a very tight virial relationship when compared with the critical dark matter mass $M_{200}$ of the form:
        \begin{equation}
    M_{200} = 10^{15} h(z)^{-1}\left ( \frac{\sigma_{DM}(M_{200},z)}{\sigma_{DM,15}} \right )^{\alpha}
        \label{eq:Evrard_rel}
        \end{equation}
where $\sigma_{DM}$ is the 1D velocity dispersion of the dark matter, or more precisely the 3D velocity dispersion divided by the $\sqrt{3}$.  For the Millennium Simulation, we use the normalization $\sigma_{DM,15} = 1093.0$ km/s and slope $\alpha = 2.94$. \citet{Evrard08} find these to be consistent between different cosmological simulations. 

The scatter in $\sigma_{1D, DM}$ at fixed $M_{200}$ from the N-body simulations for equation \ref{eq:Evrard_rel} is $\sim 5\%$. With slope $\alpha$, this inherent scatter implies a mass scatter from the virial relation to be $\sim 15\%$. However, this amount of scatter only applies when the 3D velocity dispersion is known, which is never the case in the observed Universe. \citet{Saro12} show that the real scatter in mass as measured from the line-of-sight velocity dispersion is closer to $\sim 40\%$ when 100 red-sequence galaxies are used. This huge increase in scatter stems mostly from the projection of the galaxies and their velocities along the line-of-sight.
    
\citet{Saro12} use the \citet{Delucia07} semi-analytic galaxy catalogs but they do not use the \citet{Evrard08} calibration. Instead, they re-calibrate the slope and amplitude of equation \ref{eq:Evrard_rel} using the 3-dimensional information of the {\it galaxies} as opposed to the particles. \citet{Saro12} also calibrate their measured scatter in velocity and mass against a specific set of 100 red-sequence galaxies within the 3D virial radius of each halo. In the former case, the slope and amplitude will differ from the dark matter if the galaxies are dynamically biased with respect to the dark matter. In the latter case, relationships for the scatter (e.g., as a function of the number of observed galaxies) become meaningless for larger samples (and it is quite common to have clusters with more than 100 observed galaxies in their $r-v$ phase space). 

In this work, we use the virial-mass relationship from \citet{Evrard08}. This is an important distinction, because the different semi-analytic galaxy catalogs can each have inherent dynamical biases between the dark matter and the galaxies that we would otherwise not detect. Similarly, the scatter in the measured line-of-sight velocity dispersions are determined against the particle velocity dispersions. This is also important, in that the observed line-of-sight scatter in the velocity dispersions and masses can be compared between the different semi-analytic galaxy catalogs.

    \subsection{The Halos and the Semi-analytic Galaxy Catalogs}
    \label{sec:methods_data}
    
    We select 100 halos from the Millennium Simulation to perform this study. While the halos are not chosen on the basis of any specific physical characteristics, our sample aims for fairly even mass sampling over $\sim 10^{14} - 10^{15} M_{\odot}$. The average mass $\langle M \rangle = 2.34 \times 10^{14} M_{\odot}$ and the average critical radius $\langle r_{200} \rangle = 0.95$Mpc. We then use four semi-analytic catalogs \citep{Guo11,Delucia07,Bertone07,Bower06} created using the Millennium Simulation along with the identified subhalos within each halo to test the caustic technique in section \ref{sec:results}. 
    
    \citet{Bower06} suggest that in its most basic form, a theory of galaxy formation is a set of rules motivated by physical processes which transforms a halo mass function into an observed galaxy luminosity function. In fact, most semi-analytic techniques (which define the rule-set) judge their success primarily by comparing to published galaxy luminosity functions. In this work, we are equally concerned with how well those galaxies trace the underlying radial velocity phase-space, which we use to measure the escape velocity of halos and their gravitational potentials.

Galaxies in the semi-analytic algorithms are first identified at the location of collapsed sub-structure within halos. These are nominally the sub-halos and algorithms like SUBFIND \citep{Springel01} which have been shown to work well to identify all of the sub-structure in N-body simulations (see also \citet{Knebe11}). Once identified, these sub-halos trace the positions and velocities of the galaxy population. Rules are put in place to define when and how a burst of activity (e.g. star-formation/nuclear) occurs. These sub-halos can grow in total mass (and luminosity) by accreting gas (or a model for gas), other particles, and other sub-halos through merging. It is the merging and other dynamical interactions which are responsible for altering the radial velocity phase-space and thus affecting our ability to use dynamical tracers for the halo mass.

At any given redshift (or snapshot output of an N-body simulation), the sub-halos which have survived a merger up to that point would not be good tracers of the halo radial velocity phase-space (see also Figure \ref{fig:MvsM}). This is because at the resolution of the Millennium Simulation, the sub-halos in simulations are easier to destroy compared to galaxies in the real Universe, which are much more compact and gravitationally bound. So while an interaction might destroy a sub-halo, there is no reason to think it would destroy a galaxy. In modern semi-analytic techniques, decisions (or a rule set) must be defined to decide what to do with a ``galaxy'' after its sub-halo is no longer identified in a snapshot. The most common approach is to identify the most-bound particle in the sub-halo before it was destroyed and follow it as a surviving galaxy through future snapshots (i.e., to lower redshifts). These are sometimes called ``orphans'' as they are semi-analytic galaxies that have lost their dark matter halos.

The above rule does not happen in reality: the evolution of the position and velocity of a galaxy cannot be determined from a single particle. So a technique is applied to define statistically how long a galaxy might survive before it is ripped apart in a merger (i.e., the merging time due to dynamical friction):
\begin{equation}
t_{merger} = \alpha_{fric} \frac{V_{circ}r_{SH}^2}{GM_{SH}\ln \left (1+\frac{M_{halo}}{m_{SH}}\right )}
\label{eq:time_merger}
\end{equation}
where $V_{circ}$ is the circular velocity for a mass in a gravitational potential defined by an isothermal sphere at radius $r_{SH}$ of the sub-halo and the masses of the halo and the sub-halo are $M_{halo}$ and $M_{SH}$ respectively \citep{Binney87}, and $\alpha_{fric}$ is a coefficient needed to reproduce observed luminosity functions at the luminous end \citep{Guo11}. In this rule-set, a clock is started when a galaxy's sub-halo is destroyed. The surviving galaxy is merged with its nearest galaxy or sub-halo after this time-scale has expired. When the final merging happens, the galaxy is destroyed and its stars, gas, and dark matter are distributed in various ways (i.e., it is a rule-set in the semi-analytic algorithm). However, this statistical rule-set only defines {\it when} a dark matter orphaned galaxy is destroyed, not how. For instance, the particle which represents the galaxy is no longer actually merging or changing its orbit due to dynamical interactions outside the normal particle-particle interactions. This would require a new rule-set.

\begin{figure*}
 \plotone{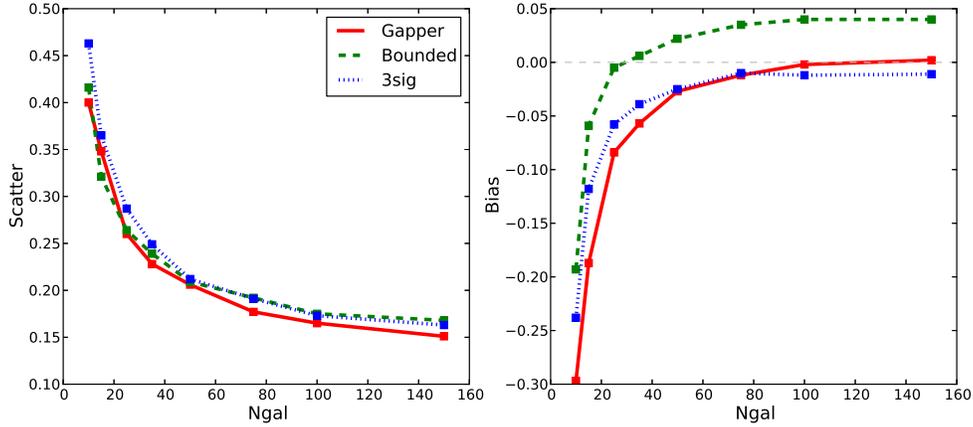}
 \caption{Log scatter ({\bf Left}) and log bias ({\bf Right}) in velocity dispersion as a function of N$_{gal}$ with interloper removal via a bounded + gapper technique (solid-red), bounded (dashed-green), and bounded + 3$\sigma$ (dotted-blue). The gapper technique both minimizes the bias and scatter of the 3 methods tested in this work for N$_{gal} > 50$.
\label{fig:interlop}}
\end{figure*}

The above description is complicated and there are numerous ways in which a semi-analytic techniques can implement dynamical events in the lifetime of a simulated galaxy. This is why we will investigate the effects on the caustic masses from different implementations of halo dynamics in the semi-analytic galaxy catalogs. 

In this analysis we start with all of the semi-analytic galaxies within a 60h$^{-1}$Mpc-length 3-dimensional box around each halo center. We then sub-select the N brightest galaxies (where N = 100,50,25, etc) within a projected radius in these volumes to create the halo radius/velocity phase-space diagrams along one or more lines-of-sight. These volumes place limits on the projected phase-space velocities that are $\pm 3000$km/s relative to the halo velocity centroids. Since the typical escape velocities are $\sim 1500$km/s, these volumes are large enough to incorporate realistic projection effects (see Figure \ref{fig:caustics}). With projection, interlopers (non-member foreground/background galaxies) can play a large role in affect both the measured phase-space density as well as the line-of-sight velocity dispersion. Therefore, an effort must be made to systematically remove the interlopers from each line-of-sight projection.

\subsection{Interloper Identification and Removal}
\label{sec:interloper_removal}
\begin{figure*}
 \plotone{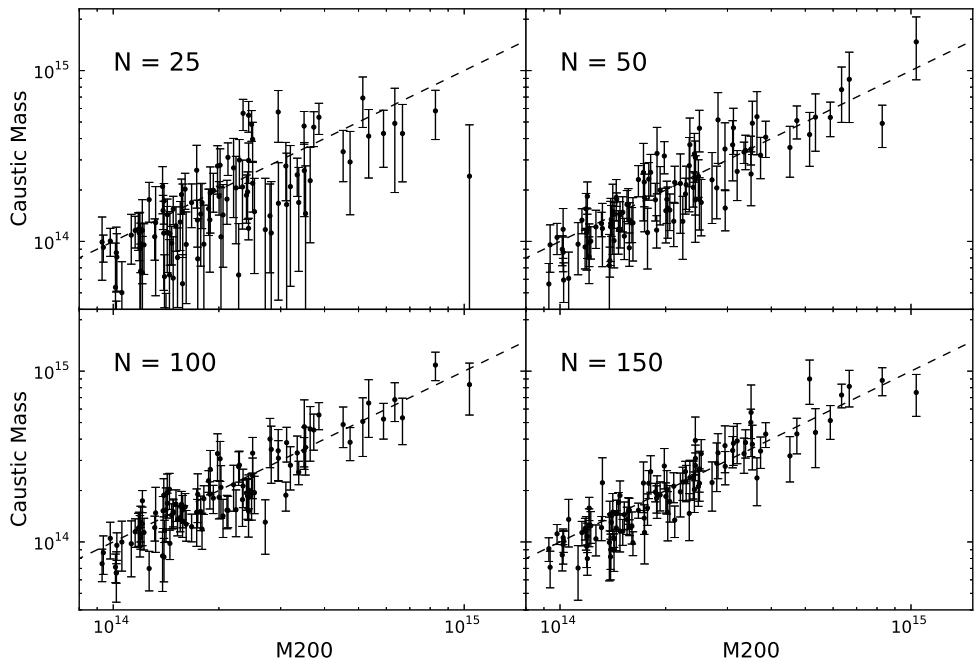}
 \caption{Inferred caustic mass for N$_{gal}$ = 25,50,100, 150 in the Guo semi-analytic sample vs $M_{200}$ for 100 halos in the Millennium Simulation. The error bars are the measured line-of-sight scatter added in quadrature with the intrinsic uncertainty in caustic mass described in  \S \ref{sec:results_los}. The dashed line is unity.
\label{fig:MvsM}}
\end{figure*}
Many methods have been devised to identify and remove interloper galaxies from a sample around a halo. These include color-based \citep{Miller05}, $\sigma$-clipping, gapper \citep{Fadda96}, and phase-space selection methods \citep{Serra13}. In this study, we do not try and optimize any particular method, but rather choose one in existence that returns a minimal bias for velocity dispersion and mass at large sampling. We limit ourselves by testing only a few basic interloper removal techniques, and compare their relative biases and resulting scatter.

In Figure \ref{fig:interlop} we compare three common techniques to measure velocity dispersions against the underlying dark matter. The first technique is to apply a simple upper/lower bound in velocity space ($\pm 3500$ km/s). The second is to iteratively remove outliers via sigma clipping ($3.5\sigma$). The third is called a shifting-gapper, where we work in the full phase-space and identify velocity gaps as a function of radius as indicators of interloping sub-structure. This shifting-gapper technique is similar to what is applied in \citet{Wing11}. Galaxies are sorted into bins as a function of radius while keeping the number in each bin constant at $N_{bin}=25$. When odd multiples of 25 are used, the last radial bin may have less than 25 galaxies. In each bin, the galaxies are sorted by their peculiar velocity and a ``f-pseudosigma" \citep{Beers90} is calculated and used as the velocity gap to remove perceived interlopers. The calculation of f-pseudosigma is an iterative process that we stop if less than 5 galaxies remain in a bin or if f-pseudosigma $< 500$ km/s. In all cases, the final interloper-cleaned velocity dispersion is calculated using a robust bi-weighted estimator, which we confirm is always less biased than a simple standard deviation \citep{Beers90}.

When applying only a simple velocity boundary, the dispersions are biased high and the scatter is large (for $N_{gal} > 25$). However, both the sigma clipping and the shifting-gapper techniques do better and equally well in recovering unbiased velocity dispersions (see also \cite{Saro12}). The shifting-gapper does a better job at reducing the scatter at all values of N$_{gal}$. This is because the technique utilizes the 2-dimensional phase-space data as opposed to the one-dimensional velocity distributions. As such, the shifting-gapper can identify sub-structure in the phase-space which sigma clipping cannot.  Clusters with poor sampling $N_{gal} < 25$, have dispersions that are biased low. Throughout the rest of this work we apply the shifting-gapper technique to measure velocity dispersions.

\section{Results}
\label{sec:results}

We measure the caustic mass $M_{c}$ inside $r_{200}$ and compare directly to the spherical halo mass $M_{200}$. In Figure \ref{fig:MvsM} we show how $M_{c}$ inferred from the N$_{gal}$ brightest galaxies inside $r_{200}$ scales with $M_{200}$. In each panel, we highlight the $M_c$ inferred masses using the Guo semi-analytic galaxies (dots). The error bars are detailed in \S \ref{sec:results_surface} \& \ref{sec:results_los}. We note that in this figure, $M_{c}$ for each of the 100 halos is measured along a single 60h$^{-1}$Mpc line-of-sight, thus representing a realistic observing scenario.

We measure the percent scatter in $\ln M_{c}$ at fixed $\ln M_{200}$ as well as the bias determined from the error weighted mean difference between $\ln M_{c}$ and $\ln M_{200}$. In order to compare to the virial masses, we also quantify the log scatter and bias of the projected line-of-sight velocity dispersion and the virial mass (equation \ref{eq:Evrard_rel}). At small N$_{gal} = 25$, the caustic masses are biased low by $\sim 15\%$ on average, and the log-normal scatter about the fit to the data is $\sim 50\%$. At N$_{gal} > 50$, the bias disappears and the scatter reduces to $\sim 30\%$ at N$_{gal} = 150$. We show a summary of the caustic mass bias and scatter in Figures \ref{fig:scatt_vs_n} and \ref{fig:bias_vs_n}. These data are presented in Tables 1-5. At fixed N$_{gal}$, the biases of the different semi-analytics agree to within their errors and so unless otherwise noted, we focus the rest of our analyses on the \citet{Guo11} galaxies. 

\subsection{The Observed Phase Space Density}
\label{sec:results_surface}

Typical observations of galaxy clusters can detect or easily measure the spectroscopic redshifts of a handful to several hundred galaxies within the virial radius of a system. Fundamental to the caustic technique is the projection of observed galaxies into a radial and velocity phase space, within which we identify a critical iso-density transition or caustic. In \S \ref{Methods}, we describe how we use a multi-dimensional gaussian kernel to estimate the underlying phase space density based on an observed tracer population. Density estimates always carry some degree of uncertainty due to limited sampling. We therefore expect that small or sparsely observed clusters will exhibit a large amount of uncertainty in the density estimation process. We quantify this uncertainty and it's effect on the caustic mass estimate by using the jack-knife technique to re-sample the phase-space. 

\begin{figure*}
\plottwo{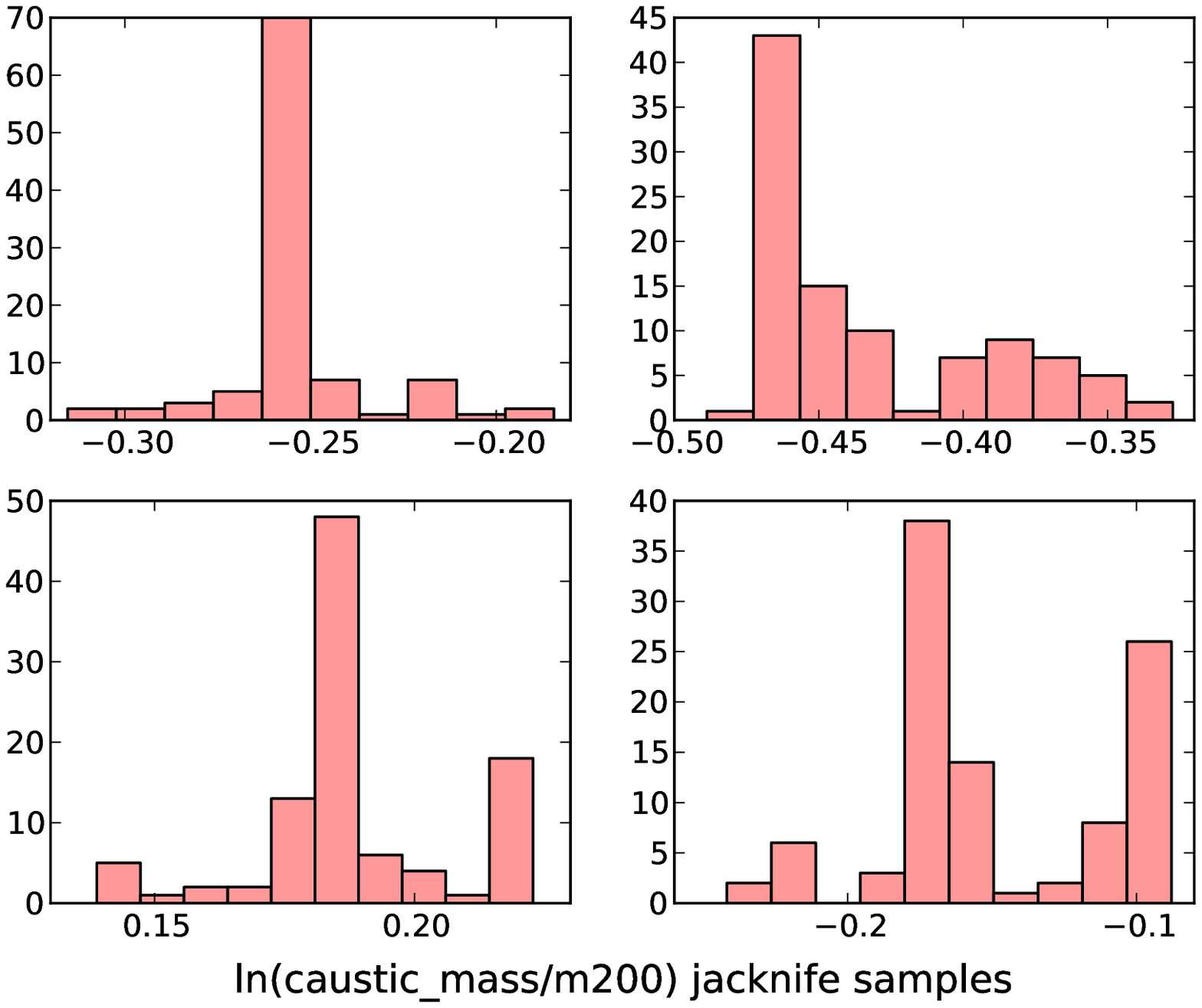}{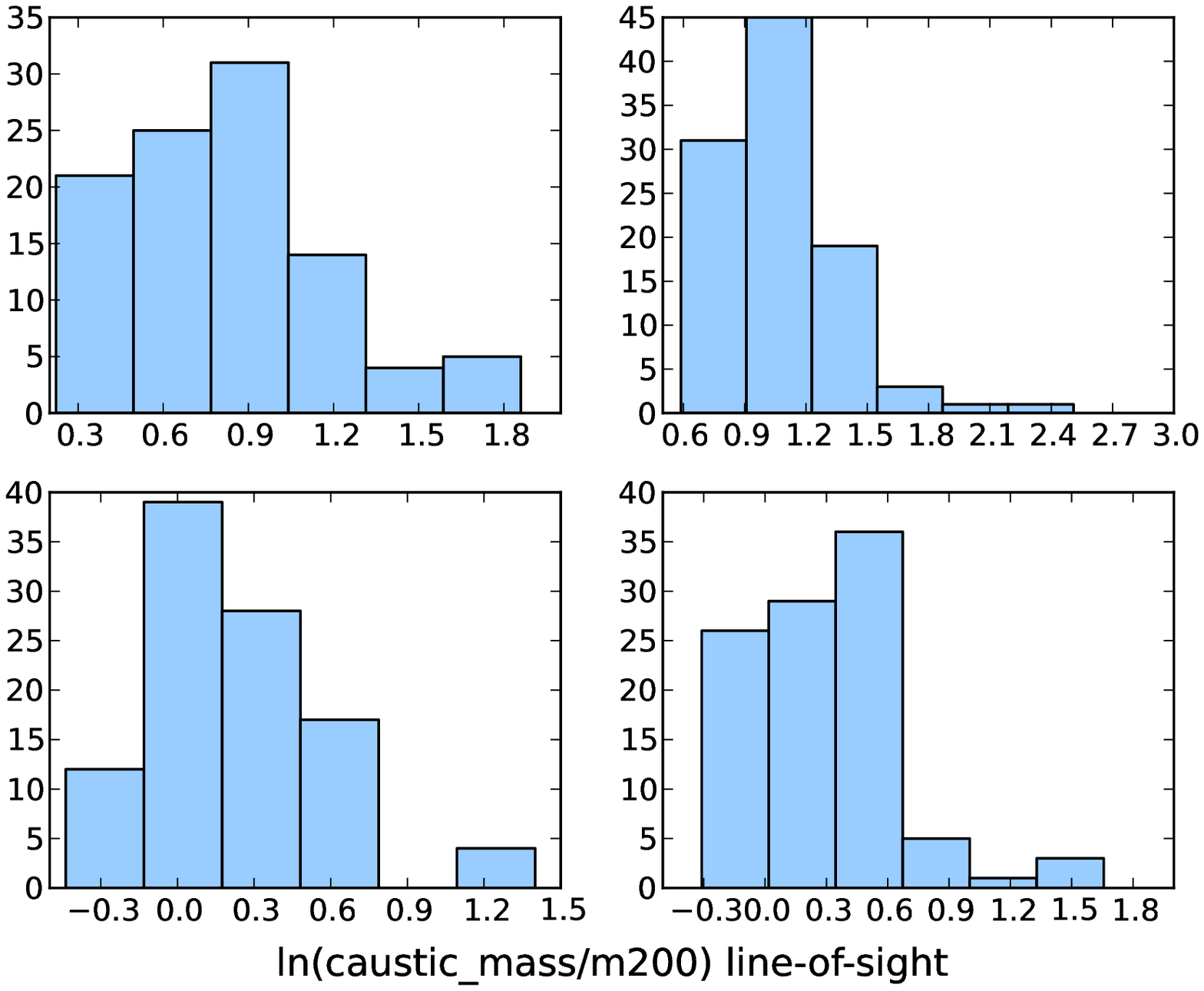}
\plottwo{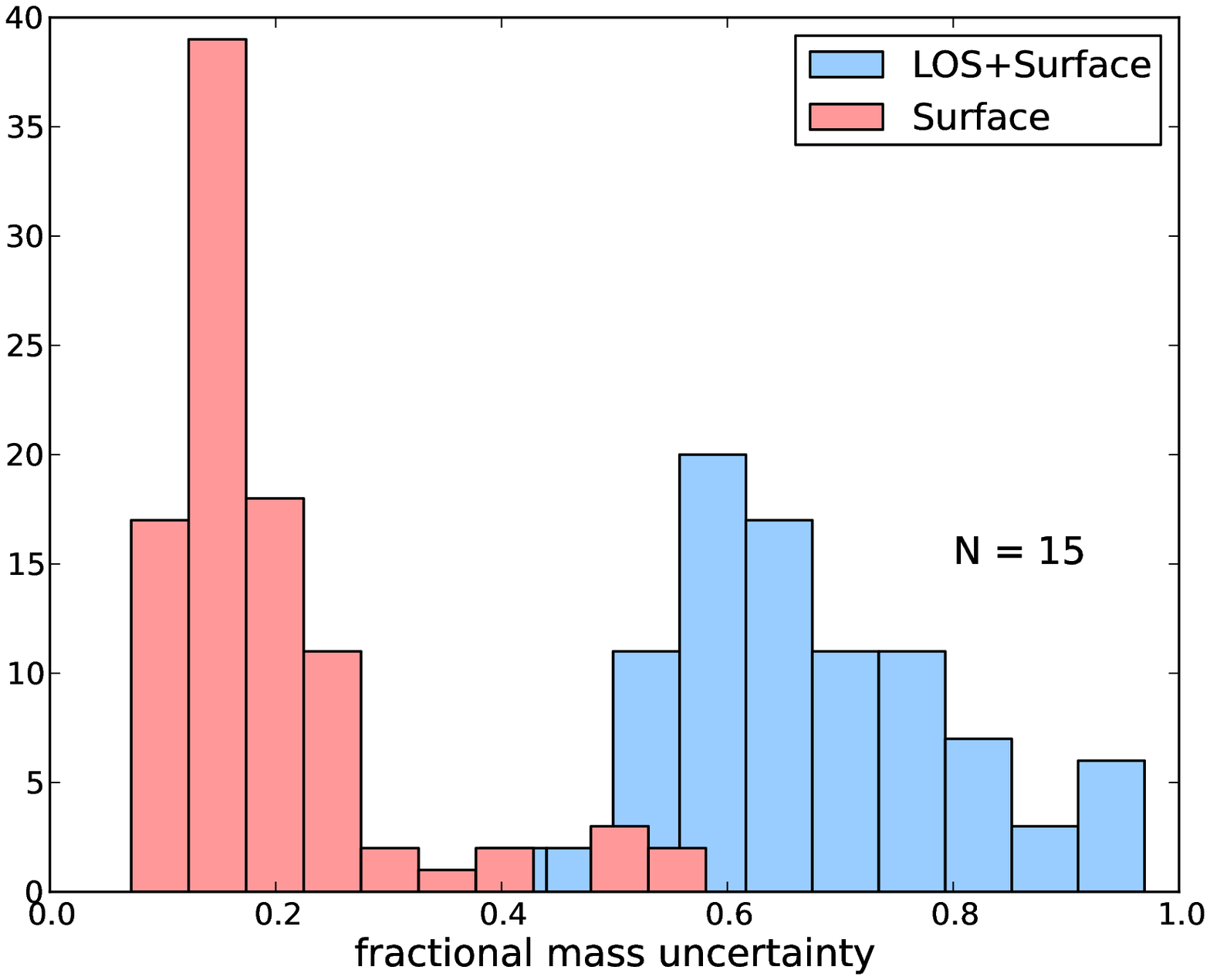}{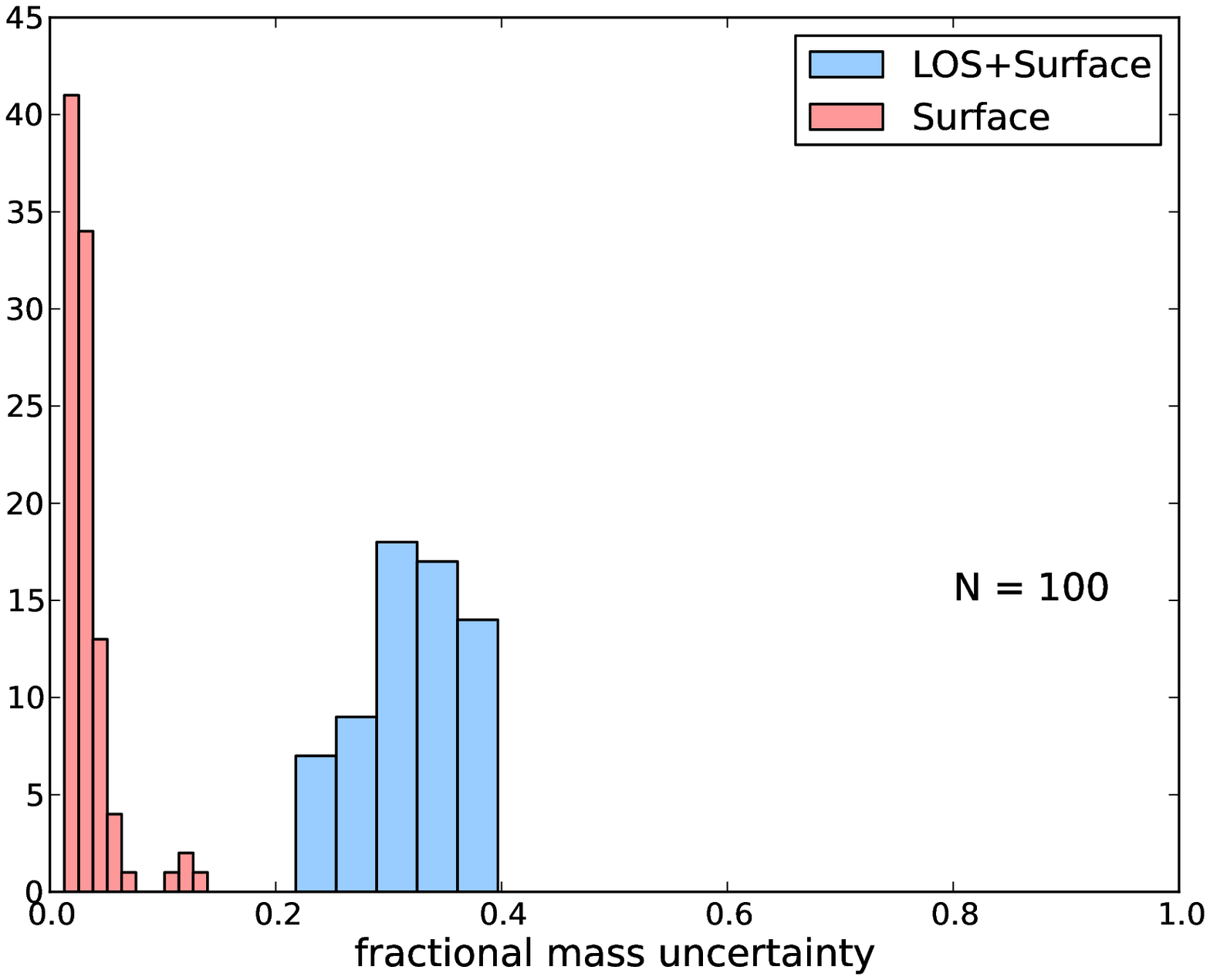}
\caption{{\bf Upper Left}: The distribution of the caustic mass compared to the true mass for four randomly chosen halos from jack-knife re-sampling of 100 galaxies along a single line-of-sight to each halo. While any given caustic mass is biased (high or low), the distributions are typically sharply peaked with small variance. This indicates that any single iso-density surface is well-defined {\bf Upper Right}: The distributions of the caustic mass compared to the true mass over 100 lines-of-sight for four other halos. These distributions are significantly wider with larger variance compared to those in the upper left panel. {\bf Lower Panels}: Histograms of the standard deviations of the distributions in the upper panels for all 100 halos; Left: N$_{gal}$ = 15 and Right: N$_{gal}$ = 100. Line-of-sight variations (blue) dominate over caustic surface uncertainties (red).\label{fig:frac_scat}}
\end{figure*}

For each jack-knife re-sampling $j$ of a cluster with N$_{gal}$, we re-calculate the velocity dispersion ($\sigma^v_j$) and re-calibrate the escape velocity iso-density contour according to equation \ref{eq:vesc_sigmav_proj} to infer a jack-knife caustic mass ($M_{c,j}$) according to equation \ref{eq:caustic_final}. We show the distribution of $\ln (M_{c,j}/M_{200})$ for four randomly chosen halos in Figure \ref{fig:frac_scat} (upper Left). For each halo, we then calculate the standard deviations of these caustic masses over the $j$ jack-knife re-samplings. We plot a histogram (red) of these standard deviations (for all 100 halos) in the lower panels of Figure \ref{fig:frac_scat} for two different values of N$_{gal}$ (15 and 100). The means of the red histograms in the lower panels of Figure \ref{fig:frac_scat} define the {\it average caustic-surface induced scatter} as a function of  N$_{gal}$.

For  N$_{gal} = 100$, the distribution of the averages of the mass uncertainties is sharply peaked around a few percent. Therefore, the surface uncertainty plays a negligible role in the total caustic mass uncertainty at N$_{gal} = 100$. By N$_{gal} = 15$, the mean of the distribution shifts to 20\% and the distribution becomes less sharply peaked. As a result of the lower density of sampling, the surface we estimate becomes less robust.

\subsection{Line-of-Sight Projections}
\label{sec:results_los}
Observationally, we only have one line-of-sight (hereafter abbreviated as ``l.o.s'') to a cluster in our universe. If clusters are not spherically symmetric in either physical or velocity space, the mass we measure will be dependent on our l.o.s to the cluster. This dependence introduces a l.o.s scatter into our observations which contributes to the scatter in mass relationships. In simulations, we are not bounded to one l.o.s and can make mass measurements along many l.o.s to a single cluster.

We show the distribution of $\ln (M_{c,los}/M_{200})$ for four randomly chosen halos in Figure \ref{fig:frac_scat} (upper Right) for 100 l.o.s. Recall that the histograms in the upper left panel of \ref{fig:frac_scat} are the 100 jack-knife re-samplings of a single line-of-sight. These l.o.s. distributions are much broader than the surface distributions. We again take the standard deviation ($\sqrt{\langle M_{c,los}^2 \rangle_{los}}$) as an estimate of the l.o.s scatter for each of our 100 halos.  The distribution of these \emph{scatters} is shown in the lower panels of Figure \ref{fig:frac_scat} in the blue histograms.  The means of the blue histograms in the lower panels of Figure \ref{fig:frac_scat} define the {\it average line-of-sight induced scatter} as a function of  N$_{gal}$. When N$_{gal} = 100$, we can see that our clusters all exhibit l.o.s scatter between 20-40\% with a mean of $\sim 30\%$. By N$_{gal} = 15$, the l.o.s scatter can be anywhere from 40-90\% for a given cluster with a mean of $\sim 65\%$.

\begin{figure*}
\plottwo{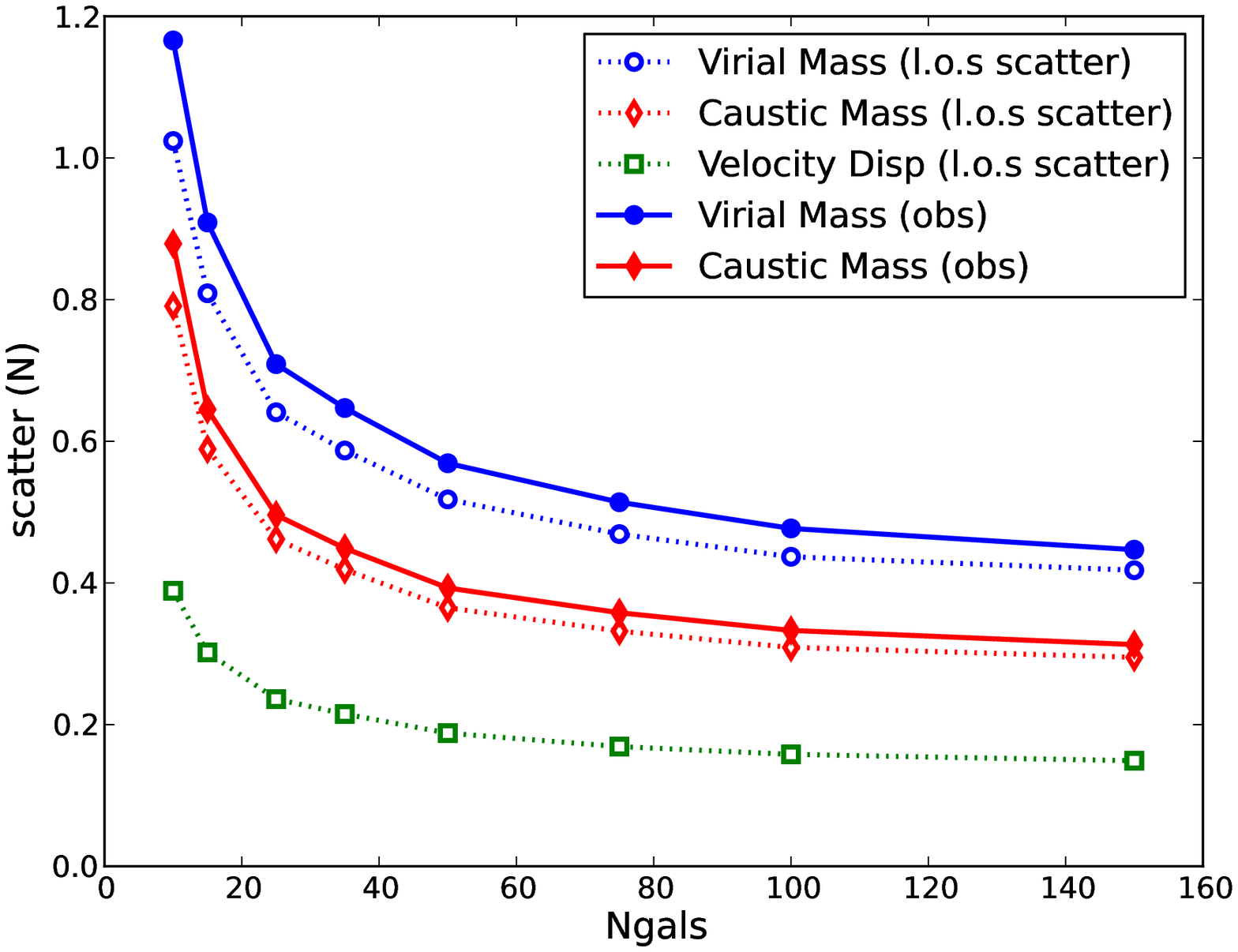}{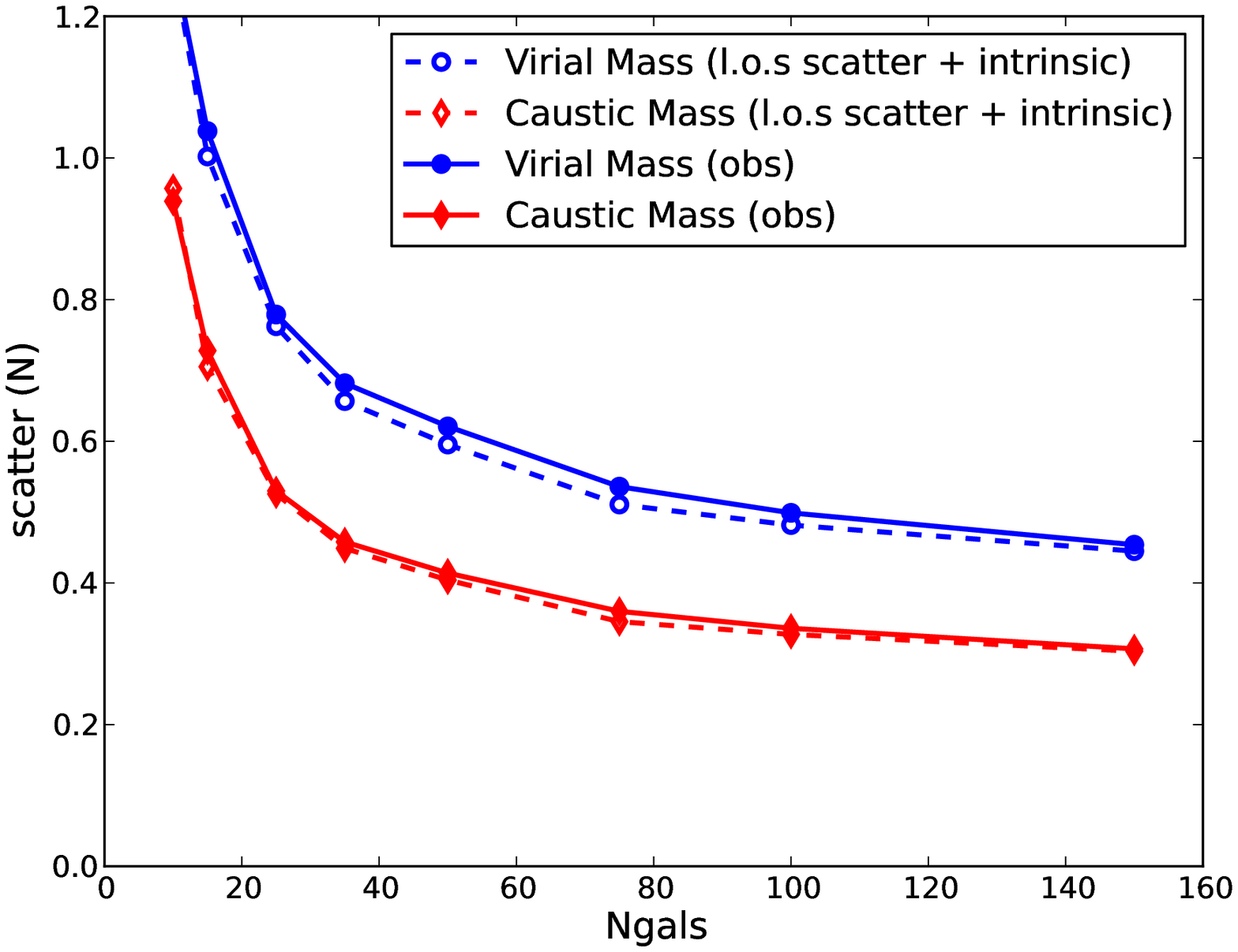}
\caption{{\bf Left}: The scatter in virial mass, caustic mass, and velocity dispersion as a function of N$_{gal}$. The solid lines represent the observed log scatter for a single line-of-sight to 100 halos from Figure \ref{fig:MvsM}. The dotted lines are the statistical representation of the line-of-sight scatter from Figure \ref{fig:frac_scat}-bottom. {\bf Right}: The solid lines are the same observed scatter (left), but are compared with the predicted mass scatters based on the summation in quadrature of the line-of-sight statistical scatters with the intrinsic 3D mass scatter.  \label{fig:scatt_vs_n}}
\end{figure*}

The mean of the l.o.s caustic mass scatter distributions are shown as a function of N$_{gal}$ in the left panel of Figure \ref{fig:scatt_vs_n} (dashed lines). Additionally, we show the mean l.o.s scatter in virial mass calculated with equation \ref{eq:Evrard_rel} and the mean l.o.s scatter in velocity dispersion.  The l.o.s scatters decrease as we increase the N$_{gal}$. The scatter in l.o.s caustic mass approaches a minimum value of $\sim 30\%$ for large N$_{gal}$ and the virial mass reaches $\sim 45\%$.  The scatter in l.o.s velocity dispersion decreases to a minimum of $\sim 15\%$ at N$_{gal} = 150$ which agrees closely with \citet{Saro12} who find a similar value using all galaxies for a large sample of halos in the \citet{Delucia07} semi-analytic catalog.

We also calculate the observed mass scatter in the caustic and virial masses directly by measuring it from Figure \ref{fig:MvsM}, which represents the scatter we would expect from a single realization of a line-of-sight to each halo. In this case, we take 100 different realizations to determine the expectation value of the scatter, which we plot as a function of $N_{gal}$ in the left panel of Figure \ref{fig:scatt_vs_n}  (solid lines). This scatter is larger than we predict from the statistical distributions determined from Figure \ref{fig:frac_scat} and shown as the dotted-lines in  \ref{fig:scatt_vs_n}. We identify the unaccounted component as originating from an intrinsic scatter in the caustic and virial masses. For instance, \citet{Evrard08} calculate the intrinsic scatter in the 3D particle velocity dispersion at fixed halo mass to be $\sim 5\%$. From Equation \ref{eq:Evrard_rel}, this translates to a 15\% intrinsic virial mass uncertainty. \citet{Gifford13b} calculate the intrinsic scatter for the caustic masses in 3D to be $\sim 10\%$, also using the particles. 

This intrinsic (i.e., 3D) uncertainty is a systematic error which should be independent of the systematic uncertainty as a result of projection effects in each l.o.s. To confirm this fact, we perform a Spearman correlation test on $M_{c,los} / \langle M_{c,los} \rangle$ vs. $M_{c,3D}/M_{200}$, where $M_{c,los}$ are all l.o.s mass measurements of a cluster, and $M_{c,3D}$ is the caustic mass measured from the 3-dimensional positions and velocities of the galaxies. We look at 100 l.o.s and find an average correlation coefficient of $0.02 \pm 0.1$ for the caustic mass and $0.03 \pm 0.1$ for the virial masses. This confirms that the two (l.o.s. and 3D) systematic uncertainties are uncorrelated. In the right panel of Figure \ref{fig:scatt_vs_n} we add the intrinsic scatter in quadrature to our statistical representations of the line-of-sight scatter (dotted lines in Figure \ref{fig:scatt_vs_n}-left). We then fully recover the observed total scatter to high precision and these are the values we present in Tables 1-5.

\begin{figure}
\plotone{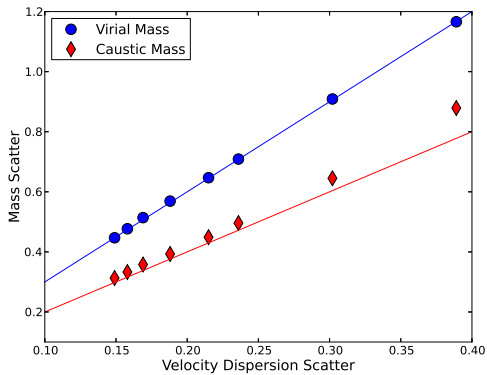}
\caption{Relationship between log scatter in velocity dispersion and log scatter in mass. The blue line predicts the virial mass uncertainties and has a slope of $2.94$ as listed in \citet{Evrard08}. The red line predicts the caustic mass uncertainties and has a slope of $2$.}
\label{fig:vdisp_vs_mass_scatt}
\end{figure}

Equation \ref{eq:Evrard_rel} implies a relationship of $\sim 3$ between the log scatter in virial mass and log scatter in velocity dispersion; however, the relationship between scatter in caustic mass and scatter in velocity dispersion is not immediately obvious as the two don't have an exact analytic relationship. We do know that the escape velocity surface is calibrated by the velocity dispersion and the caustic mass is calculated by integrating the $\langle v_{esc}^2\rangle (r)$. A naive assumption is that the scatter in velocity dispersion dominates over all other systematic scatters in the caustic technique. If so, we should find that the log scatter in caustic mass is very nearly twice that of the log scatter in velocity dispersion.

In Figure \ref{fig:vdisp_vs_mass_scatt} we show that the virial mass obeys the predicted log mass-velocity dispersion scatter relation (blue line) as expected. We also show that the sensitivity of the scatter in caustic mass is well predicted by twice the scatter in velocity dispersion (red line); however the absolute value is slightly higher due to other systematic sources of uncertainty. This implies that other forms of systematic uncertainty in the technique are indeed small, and that the line-of-sight scatter in velocity dispersion dominates the uncertainty in caustic mass, albeit to a lesser degree than it does in the virial mass.

\begin{figure*}
\plotone{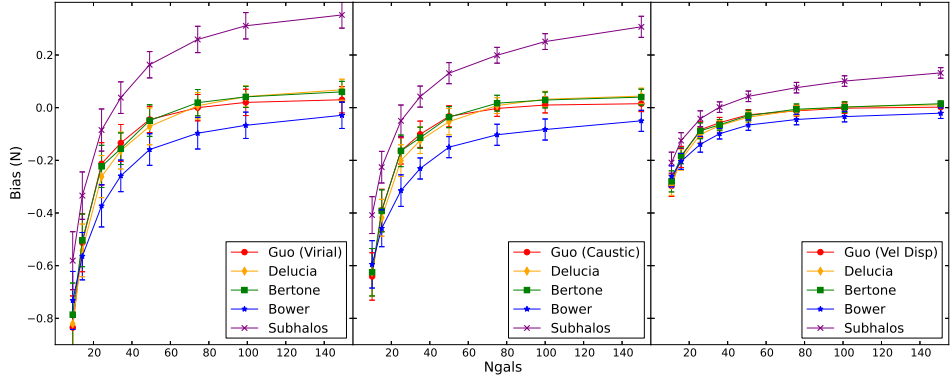}
\caption{The bias in virial mass {\bf Left}, caustic mass {\bf Middle}, and velocity dispersion {\bf Right} as a function of the number of galaxies (N$_{gal}$) used in the mass determination. The biases are shown for the Guo (red), De Lucia (orange), Bertone (green), and Bower (blue) semi-analytics as well as the subhalos (purple). N$_{gal}$ is the dominant source of bias in the caustic mass as well as the virial mass, and the semi-analytic biases all converge to within errors at high sampling.}
\label{fig:bias_vs_n}
%\epsscale{1.6}
\end{figure*}

In addition to the mass scatter for a given N$_{gal}$, we also show how the average mass bias depends on N$_{gal}$. As before with the total scatter, we measure the average bias for the 100 halos as $\langle \ln (M_{c}/M_{200}) \rangle$ after choosing only one l.o.s to each. We repeat this measurement for 100 different l.o.s and report the average sample bias. The bias in caustic mass is shown in Figure \ref{fig:bias_vs_n} (Middle) for all 4 semi-analytics and the subhalos. At small N$_{gal}$ the caustic mass can be biased very low compared with $M_{200}$. However, above N$_{gal} = 50$, the log bias is unbiased and all the semi-analytics agree to within the $1\sigma$-errors. A similar trend is seen with the virial mass (Left), but when compared with the caustic mass, the virial mass exhibits a larger bias for small N$_{gal}$. The virial mass bias, and to a lesser degree the caustic mass bias, are dependent on the velocity dispersion bias which is shown as a function of N$_{gal}$ in \ref{fig:bias_vs_n} (Right). We find that the four different semi-analytics generally agree to within the errors on the means (from the 100 l.o.s.). The \citet{Bower06} galaxies are always lower than the others, but not significantly. The sub-halos are very biased. We discuss these trends in Section \ref{sec:robust}

\subsection{Target Selection}
\label{sec:results_target}
Up to now, we have assumed that we have spectroscopic follow-up that is complete for the $N$ brightest galaxies within the projected $r_{200}$ of each halo. Here, we drop that constraint and allow more realistic targeting algorithms. We include selection based on galaxy magnitude, membership within the red-sequence, and projected distance from the cluster center. 

\begin{figure}
%\epsscale{0.8}
\plotone{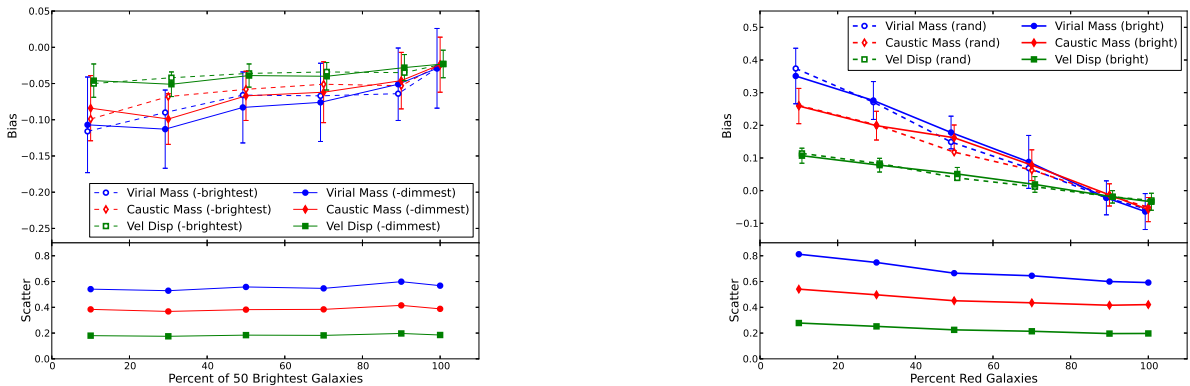}
\caption{{\bf Top}: The velocity segregation bias as a function of the fraction of the brightest galaxies within a projected $r_{200}$ used in the caustic mass. The x-axis indicates the fraction of the 50 brightest galaxies. We keep the total number of galaxies fixed by replacing bright galaxies those dimmer than the 50$^{\textrm{th}}$ brightest. We replace galaxies by starting from the dimmest (solid lines) or by starting from the brightest (dashed-see text) but always keep the 5 brightest galaxies. The errors are the uncertainties on the mean bias. {\bf Bottom}: The log sample scatter for the brighter sample (solid lines above), however the scatter for the dimmer sample is nearly identical in all cases. \label{fig:bias_vs_bright}}
%\epsscale{1.0}
\end{figure}

In Figure \ref{fig:bias_vs_bright} we keep N$_{gal}$ and color constant (i.e., only the red sequence) and show the bias (Top) and scatter (Bottom) as we decrease the fraction of original brightest (in absolute magnitude) galaxies. Starting from the sorted 50 brightest galaxies, we have different options for how we replace galaxies. For instance, we could replace them randomly, or from the brightest to the dimmest, or from the dimmest to the brightest, etc. In Figure \ref{fig:bias_vs_bright} we show two different replacement techniques starting from the \emph{dimmest} (solid lines) or  \emph{brightest} galaxies (dashed lines). We always keep the 5 brightest galaxies in each case, and we replace galaxies with those starting from the $51^{st}$ brightest within r$_{200}$. We find that the bias does not depend on in how the replacement is done. As shown in Figure \ref{fig:bias_vs_bright}, the virial mass is more affected by target selection based on galaxy luminosity than the caustic mass. To minimize brightness-induced mass biases, the \citet{Guo11} semi-analytics indicate that one should always strive to target the brightest galaxies. There is no change in the average scatter as dimmer galaxies are added to the sample.

\begin{figure}
\plotone{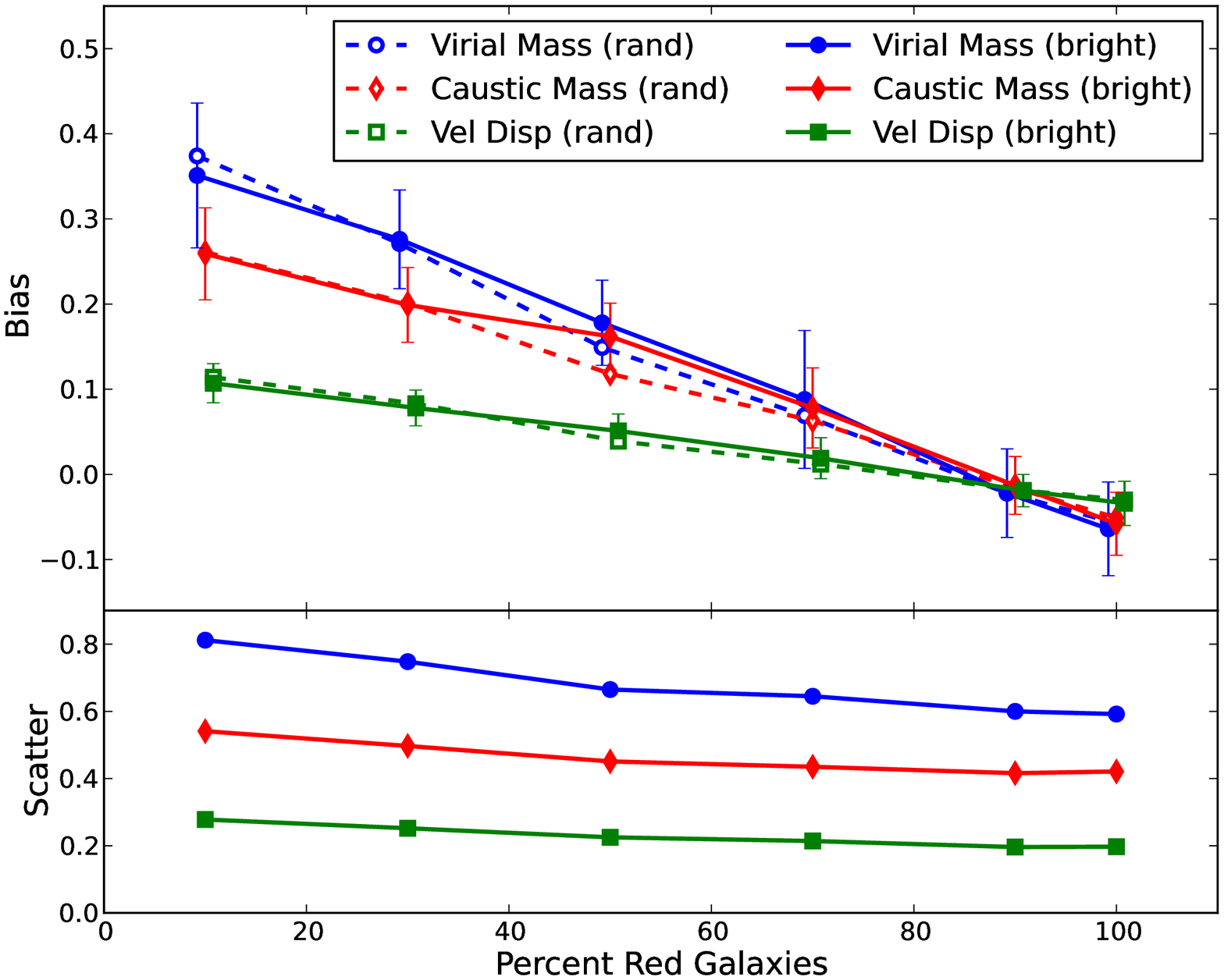}
\caption{{\bf Top}: The bias in caustic mass (red diamonds), virial mass (blue circles) and velocity dispersion (green squares) when the brightest ``x" percent of red galaxies are used out of a total of 50 (solid lines) and a random ``x" percent of red galaxies are used out of a total of 50 (dashed lines). The remaining percent added to keep N$_{gal}$ constant are the brightest blue galaxies (solid lines) and a random selection of blue galaxies (dashed lines). We detect a bias from velocity segregation based on color. {\bf Bottom}: The log sample scatter for the brighter sample (solid lines above), however the scatter for the random sample is nearly identical in all cases.}
\label{fig:red_bias_vs_red}
%\epsscale{1.0}
\end{figure}

In Figure \ref{fig:red_bias_vs_red}, we keep N$_{gal}$ constant and show the bias (Top) and scatter (Bottom) as a function of red galaxy fraction. We start with the $N = 50$ brightest red-sequence galaxies, and replace the dimmest fraction of those galaxies with the brightest blue non-red-sequence galaxies within the projected $r_{200}$. We also conduct the test by replacing the galaxies randomly, with no noticeable difference in the results. 

Color plays a bigger role in selection-induced biases and scatter when compared to brightness and again, the virial masses are more affected than the caustic masses. The velocity dispersion can be biased as much as $10\%$ higher than the baseline bias when only 25\% of the sample are bright red sequence galaxies and the rest are bright bright blue galaxies. Consequently, we see the average bias for both caustic mass and virial mass change from a slight negative bias for purely red-sequence galaxies of -5\%, to a positive bias of $\sim 25\%$ and $\sim 35\%$ respectively when we use a small fraction of the original red sequence sample. This is the color-dependent velocity segregation effect noticed in real data \citep{Carlberg97b,Goto05} and it is due to the inclusion of blue galaxies with higher infall velocities. Decreasing the fraction of bright red galaxies to bright blue galaxies can also increase the expected scatter in mass by as much as $15\%$ in caustic mass and $20\%$ in virial mass as compared with the sample of all red sequence galaxies. This is due to the higher velocities of the typically infalling blue galaxy population that can depend more on l.o.s to a cluster. It is very important to target the bright red galaxies in order to avoid this color-selection induced bias. 

\begin{figure*}
\plottwo{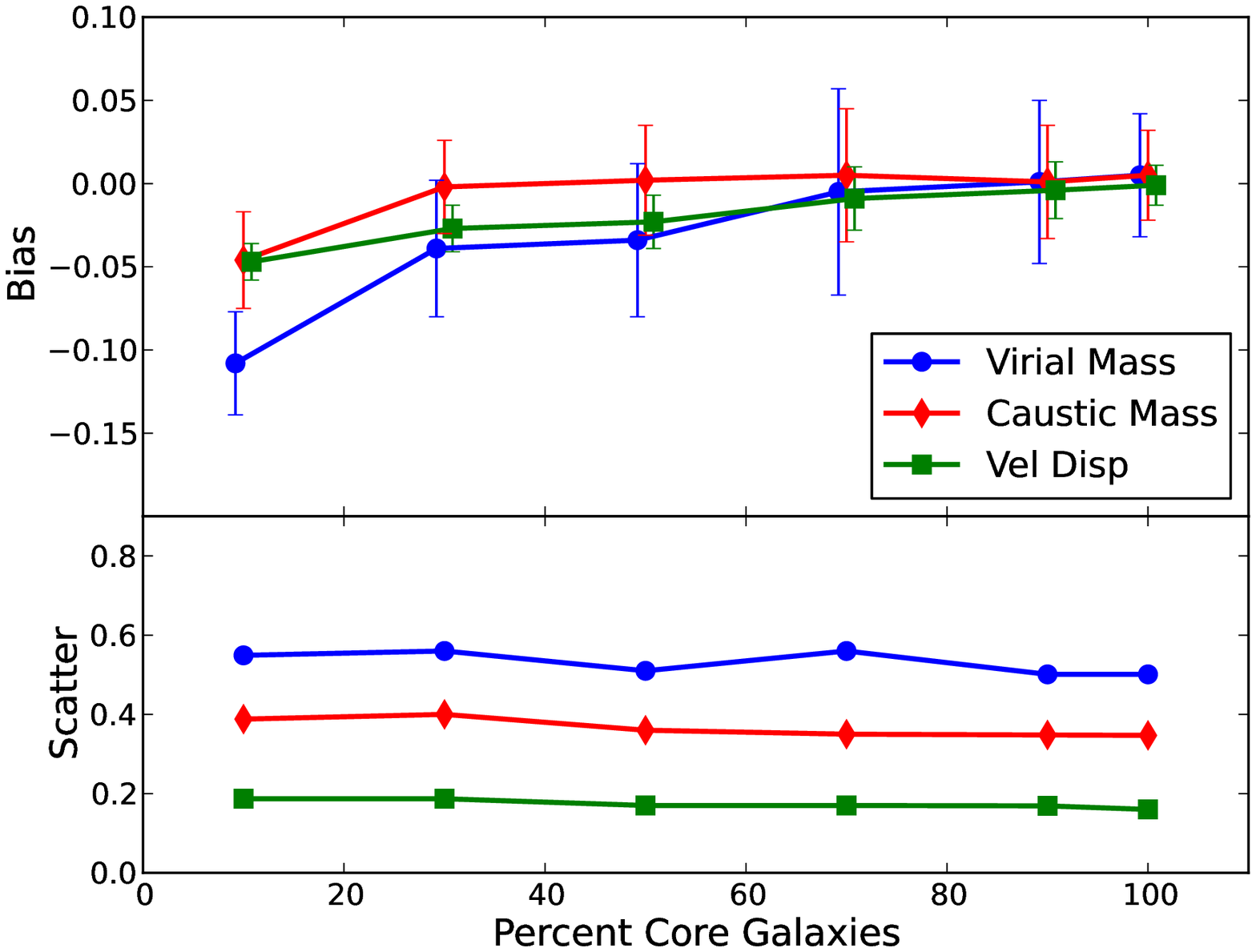}{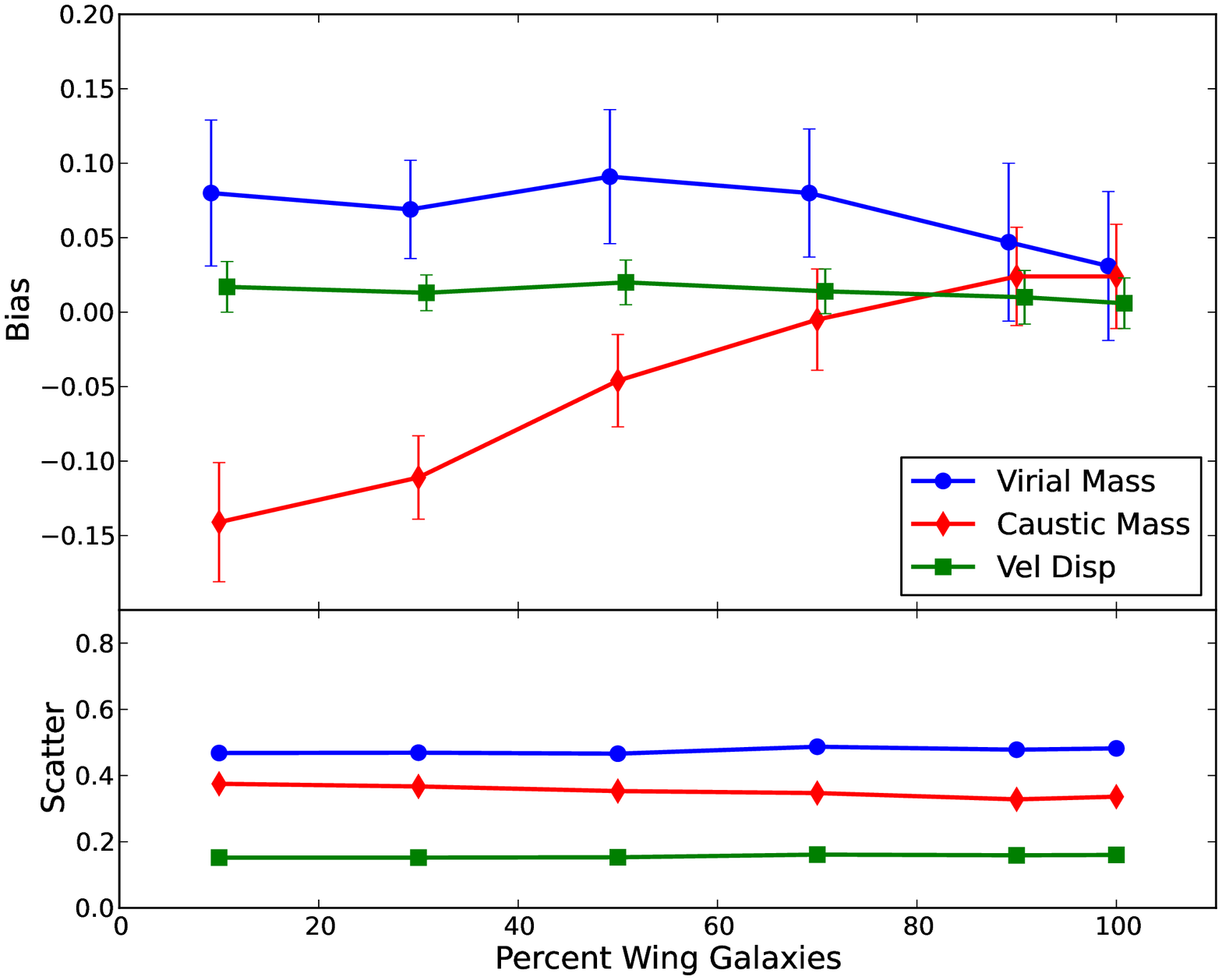}
\caption{{\bf Left}: The bias (Top) and scatter (Bottom) in caustic mass (red diamonds), virial mass (blue circles) and velocity dispersion (green squares) when the dimmest galaxies within the innermost projected 200kpc are replaced with galaxies with projected radii outside 700kpc from the center to simulate an under-sampled core. {\bf Right}: Same as left, only now the dimmest galaxies outside a projected radius of 700kpc are replaced with galaxies within the inner-most projected 200kpc to simulate an under-sampled outskirts or ``wings" region.}
\label{fig:red_bias_vs_core}
\end{figure*}

In Figure \ref{fig:red_bias_vs_core} (Left), we show the bias (Top) and scatter (Bottom) as a function of the core fraction. We start with the $N = 50$ brightest galaxies, and replace the dimmest galaxies whose projected locations are within 200kpc of the center with galaxies whose projected radii are greater than 700kpc from the center. This simulates survey data that under-samples the core due to fiber collisions or slit overlap in the dense inner regions of clusters and cannot extract spectra for all observable galaxies.  On the other hand, it is often the case where nearby massive clusters are not well-sampled out to the virial radius and beyond. In Figure \ref{fig:red_bias_vs_core} (Right) we show the bias (Top) and scatter (Bottom) as a function of the outskirt ``wing" fraction. Starting with the $N = 50$ brightest galaxies, we replace the dimmest galaxies whose projected radii are greater than 700kpc with galaxies that are within 200kpc from the center.

In the first case, the more we under-sample the core relative to the outskirts, the more biased our estimates become of mass and velocity dispersion. This is true regardless of whether we use the virial relation or the caustic technique. No matter how much we under-sample the core, the caustic inferred masses are less biased than the virial estimates. This is due to the fact that the caustic technique does not depend as heavily on uncertainty in velocity dispersion (see Figure \ref{fig:vdisp_vs_mass_scatt}), and under-sampling the projected core of a cluster will lower the measured velocity dispersion. We see no effect on the mass scatter when under-sampling the core.

When we under-sample the outskirts or ``wings" relative to the core, we see dramatically different responses from the caustic and virial mass estimators. Under-sampling the outskirts has the effect of raising the measured velocity dispersion. This is because the galaxies with large l.o.s velocity are preferentially projected near the cluster core while under-sampling the outskirts removes galaxies with preferentially smaller l.o.s velocities. This raises the virial mass estimate as it is directly proportional to velocity dispersion. However, the caustic inferred masses exhibit the opposite trend since we are reducing the phase-space density in the outer regions by removing these galaxies. The mass scatter is unaffected by under-sampling the outskirts. To minimize radial-selection induced biases, it is important to evenly sample the phase space.

\section{Discussion}
\label{sec:Discussion}

With the availability of large and complete spectroscopic surveys like the Sloan Digital Sky Survey \citep[SDSS]{Stroughton02} and GAMA (GAlaxy Mass and Assembly GAMA) \citep{Driver11}, highly multiplexed multi-object optical spectroscopy on 8m class telescopes (IMACS \citep{Dressler11}, VIMOS \citep{LeFevre03}, DEIMOS \citep{Davis03}), and the latest generation of multi-object near-infrared spectrographs (MMIRS \citep{McLeod04}, Flamingos 2 \citep{Eikenberry06}, MOSFIRE \citep{McLean08}), we can measure galaxy large numbers of galaxy velocities in clusters from $z \sim 0$ to $z \sim 1.5$. And just as we can use three different observables to infer the masses of clusters (see Section \ref{sec:intro}), we can use the galaxy velocities in three different ways to infer dynamical mass (e.g., via virial scaling, the Jean's equation, and the escape velocity). The questions then are how well do these different dynamical techniques work under realistic observing conditions and how do they compare. Here, we compare the virial relation via the velocity dispersion \citep{Evrard08} and the escape velocity mass inferred from the caustic technique \citep{Diaferio97}.

\subsection{Robustness of the Caustic Technique to the Galaxy Models}
\label{sec:robust}

We examine four different semi-analytic mock galaxy catalogs, as well as the sub-halos and the particles in the Millennium Simulation. The measured scatter and bias in the inferred dynamical masses are quantified in Tables 1-5. The level of scatter and bias varies by only a few percent between most of the different semi-analytic tracers. The exception is for the \citet{Bower06} galaxies, which sometimes differ by slightly more than the 1$\sigma$ errors on the bias estimates. In one sense, this robustness is not surprising, since the majority of galaxies in the semi-analytics are attached to sub-halos, which are the same for each of the mock galaxy catalogs. However, there are a variety of rules which are applied to the ``orphan'' population (see Section \ref{sec:methods_data}). For the satellite population, these orphan galaxies are quite common at low stellar masses, while the exact fraction depends on the semi-analytic rule-set. In the semi-analytic catalogs studied here, the fraction of $N=50$ brightest galaxies that are orphans ranges from 50\% (Bower) to 25\% (Guo).

Consider some of the differences in the orphan population between the \citet{Bower06} semi-analytics compared to \citet{Delucia07}, \citet{Bertone07}, and \citet{Guo11}: The Bower sample uses a merger tree that differed in how friend-of-friend groups were defined, how spuriously linked halos were handled, how independent halos were identified, and how descendants of halos were tracked through time; the Bower sample includes the sub-halo orbital energy and angular momentum when calculating the merging timescale of the orphans; the \citet{Guo11} semi-analytics go one step further by modeling the orbital decay of the orphans before they are destroyed. 

These algorithmic differences manifest as variations in the spatial distribution of the satellites within the halos. The De Lucia semi-analytic galaxies have much flatter density profiles within the virial radius when compared to the \citet{Bower06} galaxies \citep{Budzynski12}.  Since the $r-v$ phase-space is different between the semi-analytics, one might expect that the caustic technique (which is based on the phase-space density) would also differ significantly. However, this is not the case and even when the orphan fraction and radial densities differ by factors of two or more, the caustic masses vary by $< 10\%$. We interpret this lack of difference between the different semi-analytic techniques as a measure of the robustness of the technique to large systematic variations in the $r-v$ velocity distribution function.

The one exception to this robustness is for the sub-halo population where we detect large biases in the velocity dispersion compared to the dark matter. Since the caustic mass is calibrated using the velocity dispersion, this bias carries through into a bias in both the virial mass and the caustic mass. We attribute this to how poorly the sub-halos trace the $r-v$ phase-space within halos in simulations with resolution similar to the Millennium Simulation (see also \citet{Budzynski12}). 

One could ask whether the semi-analytic techniques have converged in their representation of the $r-v$ phase-space inside clusters, as represented by the fraction of orphan galaxies. \citet{Faltenbacher06} have shown that at higher resolutions, galaxies are no longer orphaned from sub-halos. Instead, a new ``crisis'' of sub-halo over-abundance presents itself. \citet{Faltenbacher06} also show that this crisis is averted through rule-sets analogous to the ones used at lower resolution. In other words, instead of following the most-bound particle of a destroyed sub-halo, one attaches semi-analytic galaxies only to those sub-halos that previously achieved some minimal mass threshold before entering the cluster. The velocity dispersion of these semi-analytic galaxies is unbiased with respect to the dark matter, which is what we already find here.

\subsection{The Scatter and Bias of Dynamical Cluster Mass Estimates}
While none of these semi-analytic prescriptions is how nature places galaxies into halos, we have established that the virial relation and the caustic technique are robust to variations in their rule-sets. This baseline allows for a comparison of the absolute levels of the inferred mass biases and scatters between the virial and the caustic techniques.

For a fixed number of tracers projected to lie within $R_{200}$ of the clusters, the scatter in the caustic mass $\ln M_{c}$ at fixed $\ln M_{200}$ is about 1/3 smaller than $M_{virial}$.  In other words, at fixed N, the caustic technique is a more precise estimate of the true halo mass compared to the virial relation. The growth of the mass scatter as $N_{gal}$ decreases is the same for the two techniques. As with the virial relation and the velocity dispersion, the dominant component of the scatter in the caustic mass is from the line-of-sight variations in the observed velocity dispersion and phase-space density (see also \citet{White10}; \citet{Saro12}). Uncertainties in $M_{c}$ induced from the sampling of the phase space do not contribute unless N$_{gal} < 25$, below which they quickly grow to be a dominant component. 

In Figure \ref{fig:bias_vs_n} we show the strong dependence of the bias on the number of tracers used in the mass calculation. \citet{Saro12} suggest that this was due to dynamical friction, since we (as they) explicitly restrict the sub-samples to the brightest galaxies. However at fixed N, large variations in the brightness distribution of the galaxies has only a small affect on the observed bias. The same can be said for color selection (see Figure \ref{fig:red_bias_vs_red}). Thus while velocity segregation from sub-sampling plays a role, it is minimal compared to the number of galaxies used in the mass determination. This holds for both the virial masses as well as the caustic masses. Put simply, if a targeting algorithm requires a trade-off between number and color, number wins.

Regardless of how biases are introduced into the velocity distributions (e.g., through number, brightness, or color) the caustic mass is less affected than the virial mass. The trends in the biases for the virial and caustics masses are similar. The exception is when radial selection constraints are applied. We find that the velocity dispersion is more susceptible to incompletenesses in the core sampling while the caustic technique is more susceptible to incompletenesses in the cluster outskirts. The former trend can be understood from a dynamically cold population of galaxies which dominates the dispersion measurement at low sampling. The latter is also a sampling issue, where the iso-density contour gets lost in the background for low sampling in the cluster outskirts.

\subsection{The Calibration Factor}
Finally, we note that the implementation of the caustic technique utilizes a step which calibrates to simulations (see $\mathcal{F}_{\beta}$ in equation \ref{eq:caustic_final}). This is no different than the virial calibration (equation \ref{eq:Evrard_rel}). \citet{Diaferio99} find $\langle \mathcal{F}_{\beta}(r) \rangle = 0.5$ and \citet{Serra11} find $\langle \mathcal{F}_{\beta}(r) \rangle = 0.7$, which corresponds to a systematic uncertainty of 30\% in mass. This is larger than any biases we detect from sample selection.

One explanation the difference in the value for $\mathcal{F}_{\beta}$ is in how \citet{Diaferio99} and  \citet{Serra11} define their mock galaxy catalogs: the former used only halos (with ten particles or more) while the latter used only particles to define galaxies. Since sub-halos in N-body simulations are biased tracers of the dark matter velocity dispersions, the calibration factor and caustic masses will also be biased (see Figure \ref{fig:bias_vs_n} middle and right panels). We test this hypothesis by re-calibrating  $\mathcal{F}_{\beta}$ using the Millennium Simulation sub-halos alone. The high halo velocity dispersions of the sub-halos require a lower calibration factor than from the dark matter. We find $\langle \mathcal{F}_{\beta}(r) \rangle = 0.5$, identical to what is used in  \citet{Diaferio99}, who note that their own halo velocity dispersions are biased with respect to the dark matter in their simulations. \citet{Serra11} explicitly used dark matter particle positions and velocities to define their galaxies and calibrate $\mathcal{F}_{\beta}$. The semi-analytics we use lie in between these two extremes, which explains why we find $\langle \mathcal{F}_{\beta}(r) \rangle = 0.65$. 

\citet{Serra11} notes that  $\mathcal{F}_{\beta}$ is not constant with radius in simulations and that the caustic mass profiles can over-estimate the mass within $\sim 0.3r_{200}$, where velocity anisotropies become smaller. Likewise, \citet{Geller13} compare caustic-derived and weak-lensing-derived mass profiles and find disagreement within $0.5r_{200}$, as expected from the anisotropy profiles in simulations. In this work, we are not measuring mass profiles but instead the integrated masses of the clusters out to $r_{200}$. As shown in \citet{Serra11}, unbiased halo masses require an appropriate measure of $\langle \mathcal{F}_{\beta}(r) \rangle = \int_{0}^{r} \mathcal{F}_{\beta}(r) dx/r$ from Equation \ref{eq:caustic_final}, which is what we do to achieve unbiased caustic masses for the \citet{Guo11} semi-analytic galaxies. This is without question a tuning step which carries with it a certain level of additional systematic uncertainty.

It is very unlikely that the radius-velocity phase-space of galaxies in the Universe is represented by the sub-halo population in N-body simulations (regardless of resolution). Current observational constraints on the bias between the galaxy and dark matter velocity dispersions are measured indirectly (e.g., \citet{Rines08}) and find it to be $< 5\%$ when a multitude of independent cosmological priors are leveraged. From the current simulation and observational work, it is not likely that $\langle \mathcal{F}_{\beta}(r) \rangle = 0.5$, which corresponds to a galaxy/dark matter velocity bias of 10\% in a $\Lambda$CDM simulation (see Table 2 sub-halos). If galaxies and dark matter have dispersions that are unbiased with respect to each other, the caustic masses (integrated to $r_{200}$) would be low by $\sim$20\% using $\langle \mathcal{F}_{\beta}(r) \rangle = 0.5$ \citep{Geller13,Rines13}. 

Ideally, the simulation calibration factor can be dropped entirely, and in \citet{Gifford13b} we present a revised derivation of the caustic technique that relies only on the observable parameters: the NFW density scale parameter, $\langle v^2_{esc,los} \rangle$, and $\langle \beta \rangle$ and their statistical and systematic uncertainties.

\section{Summary}
In this work, our main conclusions are as follows.

\begin{itemize}
\item We measure velocity dispersions, virial masses ($M_{virial}$), and caustic masses ($M_{c}$) for 100 halos in the Millennium Simulation with masses $10^{14} - 10^{15}M_{\odot}$. These halos exist at low redshift ($z < 0.15$) and are populated with galaxies via four different semi-analytic prescriptions \citep{Guo11,Delucia07,Bower06,Bertone07}.
\item The resulting scatter and bias in $M_c$ relative to the halo mass ($M_{200}$) is robust and largely independent of the semi-analytic prescription used to populate cluster-size dark matter halos with galaxies. The exception is when only subhalos are observed which do not trace the dark matter velocity field and measured velocity dispersions. The velocity dispersion for the subhalos can be biased high by $10$-$15\%$ resulting in a caustic mass bias of $20$-$30\%$.
\item As with the $M_{virial}$, the dominant component of the scatter in $M_c$ is from the line-of-sight variations in the observed velocity dispersion. However, for a fixed number of tracers (N$_{gal}$) projected to lie within $R_{200}$ of the clusters, the scatter in the caustic mass $\ln (M_c|M_{200})$ is $\sim 30\%$ at N$_{gal} > 50$ which is  $\sim 1/3$ smaller than $M_{virial}$.
\item The bias in $M_c$ relative to $M_{200}$ is strongly dependent on N$_{gal}$. While color selection, radial completeness, and magnitude can play a role in inducing bias depending on how a cluster is observed, their effect is much less than N$_{gal}$. Given a choice, it is better to use a full, unrestricted sample to estimate caustic masses rather than a color selected (such as red sequence) sample to achieve larger numbers.
\item We find a caustic mass calibration factor $\mathcal{F}_{\beta} = 0.65$. This differs from the calibrations based on either just the dark matter particles ($\mathcal{F}_{\beta} = 0.7$) or just the subhalos ($\mathcal{F}_{\beta} = 0.5$). While galaxy/DM velocity bias affects the caustic mass less than the dynamical mass, it is an important component of the total systematic uncertainty.
\end{itemize}

\section{Acknowledgements}
The authors made use of the FLUX High Performance Computing Cluster at the University of Michigan. The Millennium Simulation databases used in this paper and the web application providing online access to them were constructed as part of the activities of the German Astrophysical Virtual Observatory (GAVO). The authors want to especially thank Gerard Lemson for his assistance and access to the particle data. We also wish to thank the referee, Craig Harrison, August Evrard, Heidi Wu, Risa Wechsler, and Michael Busha for their helpful comments and discussion. This material is based upon work supported by the National Science Foundation Graduate Student Research Fellowship under Grant No. DGE 1256260.

\bibliographystyle{apj}
%\bibliography{gifftex}{}

\begin{table}[b]
\begin{center}
%\scalebox{0.7}{
%\label{tab:bias_scatter}
%\caption{}%\label{sample_bias&scatter_table}}
\begin{tabular}{c | cc | cc | cc | c}
\multicolumn{8}{c}{Guo} \\
\multicolumn{8}{c}{Velocity Dispersion and Caustic Mass bias/scatter} \\
 & \multicolumn{2}{c |}{${\sigma}^{v}_{N}$} & \multicolumn{2}{c}{Virial Mass} & \multicolumn{2}{c}{Caustic Mass} & Orphans\\
\hline
N & Bias & Scatter & Bias & Scatter & Bias & Scatter & Fraction\\
 & $\langle \ln(\sigma^v/\sigma_{dm})\rangle$ & $\langle \ln(\sigma^v/\sigma_{dm})^2 \rangle^{1/2}$ & $\langle \ln(M_v/M_{200})\rangle$ & $\langle \ln(M_v/M_{200})^2 \rangle^{1/2}$ & $\langle \ln(M_c/M_{200})\rangle$ & $\langle \ln(M_c/M_{200})^2 \rangle^{1/2}$ & \\
\hline
150 & 0.00$\pm$0.02 & 0.15 & 0.03$\pm$0.05 & 0.45 & 0.01$\pm$0.03 & 0.31 & 0.40\\
100 & 0.00$\pm$0.02 & 0.17 & 0.02$\pm$0.05 & 0.50 & 0.01$\pm$0.03 & 0.33 & 0.34\\
75 & -0.01$\pm$0.02 & 0.18 & 0.00$\pm$0.05 & 0.54 & 0.00$\pm$0.03 & 0.36 & 0.30\\
50 & -0.03$\pm$0.02 & 0.21 & -0.05$\pm$0.05 & 0.62 & -0.03$\pm$0.04 & 0.41 & 0.26\\
35 & -0.06$\pm$0.02 & 0.23 & -0.13$\pm$0.07 & 0.68 & -0.10$\pm$0.05 & 0.46 & 0.25\\    
25 & -0.08$\pm$0.03 & 0.26 & -0.21$\pm$0.08 & 0.78 & -0.16$\pm$0.05 & 0.53 & 0.23\\
15 & -0.19$\pm$0.04 & 0.35 & -0.51$\pm$0.11 & 1.04 & -0.39$\pm$0.08 & 0.73 & 0.21\\
10 & -0.30$\pm$0.04 & 0.40 & -0.84$\pm$0.12 & 1.28 & -0.64$\pm$0.09 & 0.94 & 0.21\\
\end{tabular}%}
\end{center}
\end{table}

\begin{table}
\begin{center}
%\scalebox{0.7}{
%\caption{}%\label{sample_bias&scatter_table}}
\begin{tabular}{c | cc | cc | cc | c}
\multicolumn{8}{c}{De Lucia} \\
\multicolumn{8}{c}{Velocity Dispersion and Caustic Mass bias/scatter} \\
 & \multicolumn{2}{c |}{${\sigma}^{v}_{N}$} & \multicolumn{2}{c}{Virial Mass} & \multicolumn{2}{c}{Caustic Mass} & Orphans\\
\hline
N & Bias & Scatter & Bias & Scatter & Bias & Scatter & Fraction\\
 & $\langle \ln(\sigma^v/\sigma_{dm})\rangle$ & $\langle \ln(\sigma^v/\sigma_{dm})^2 \rangle^{1/2}$ & $\langle \ln(M_v/M_{200})\rangle$ & $\langle \ln(M_v/M_{200})^2 \rangle^{1/2}$ & $\langle \ln(M_c/M_{200})\rangle$ & $\langle \ln(M_c/M_{200})^2 \rangle^{1/2}$ & \\
\hline
150 & 0.01$\pm$0.01 & 0.15 & 0.06$\pm$0.04 & 0.45 & 0.04$\pm$0.03 & 0.31 & 0.36\\
100 & 0.00$\pm$0.02 & 0.16 & 0.04$\pm$0.04 & 0.49 & 0.03$\pm$0.03 & 0.33 & 0.32\\
75 & -0.01$\pm$0.02 & 0.18 & 0.01$\pm$0.05 & 0.54 & 0.01$\pm$0.03 & 0.36 & 0.30\\
50 & -0.04$\pm$0.02 & 0.21 & -0.07$\pm$0.07 & 0.62 & -0.05$\pm$0.05 & 0.42 & 0.28\\
35 & -0.07$\pm$0.02 & 0.23 & -0.16$\pm$0.07 & 0.68 & -0.12$\pm$0.05 & 0.46 & 0.28\\    
25 & -0.10$\pm$0.03 & 0.26 & -0.26$\pm$0.08 & 0.80 & -0.20$\pm$0.06 & 0.54 & 0.26\\
15 & -0.20$\pm$0.04 & 0.35 & -0.54$\pm$0.10 & 1.03 & -0.42$\pm$0.07 & 0.73 & 0.25\\
10 & -0.29$\pm$0.04 & 0.43 & -0.82$\pm$0.13 & 1.29 & -0.63$\pm$0.09 & 0.96 & 0.23\\
\end{tabular}%}
\end{center}
\end{table}

\begin{table}
\begin{center}
%\scalebox{0.7}{
%\caption{}%\label{sample_bias&scatter_table}}
\begin{tabular}{c | cc | cc | cc | c}
\multicolumn{8}{c}{Bertone} \\
\multicolumn{8}{c}{Velocity Dispersion and Caustic Mass bias/scatter} \\
 & \multicolumn{2}{c |}{${\sigma}^{v}_{N}$} & \multicolumn{2}{c}{Virial Mass} & \multicolumn{2}{c}{Caustic Mass} & Orphans\\
\hline
N & Bias & Scatter & Bias & Scatter & Bias & Scatter & Fraction\\
 & $\langle \ln(\sigma^v/\sigma_{dm})\rangle$ & $\langle \ln(\sigma^v/\sigma_{dm})^2 \rangle^{1/2}$ & $\langle \ln(M_v/M_{200})\rangle$ & $\langle \ln(M_v/M_{200})^2 \rangle^{1/2}$ & $\langle \ln(M_c/M_{200})\rangle$ & $\langle \ln(M_c/M_{200})^2 \rangle^{1/2}$ & \\
\hline
150 & 0.02$\pm$0.01 & 0.15 & 0.06$\pm$0.04 & 0.44 & 0.04$\pm$0.03 & 0.29 & 0.39\\
100 & 0.00$\pm$0.02 & 0.16 & 0.04$\pm$0.04 & 0.48 & 0.03$\pm$0.03 & 0.32 & 0.34\\
75 & -0.01$\pm$0.02 & 0.18 & 0.02$\pm$0.05 & 0.54 & 0.02$\pm$0.03 & 0.36 & 0.31\\
50 & -0.03$\pm$0.02 & 0.21 & -0.05$\pm$0.06 & 0.62 & -0.04$\pm$0.04 & 0.41 & 0.27\\
35 & -0.07$\pm$0.02 & 0.23 & -0.16$\pm$0.06 & 0.68 & -0.11$\pm$0.04 & 0.46 & 0.26\\    
25 & -0.09$\pm$0.03 & 0.27 & -0.22$\pm$0.08 & 0.80 & -0.16$\pm$0.06 & 0.54 & 0.25\\
15 & -0.18$\pm$0.03 & 0.34 & -0.50$\pm$0.10 & 1.02 & -0.39$\pm$0.08 & 0.72 & 0.25\\
10 & -0.28$\pm$0.04 & 0.42 & -0.79$\pm$0.12 & 1.27 & -0.63$\pm$0.09 & 0.95 & 0.25\\
\end{tabular}%}
\end{center}
\end{table}

\begin{table}
\begin{center}
%\scalebox{0.7}{
%\caption{}%\label{sample_bias&scatter_table}}
\begin{tabular}{c | cc | cc | cc | c}
\multicolumn{8}{c}{Bower} \\
\multicolumn{8}{c}{Velocity Dispersion and Caustic Mass bias/scatter} \\
 & \multicolumn{2}{c |}{${\sigma}^{v}_{N}$} & \multicolumn{2}{c}{Virial Mass} & \multicolumn{2}{c}{Caustic Mass} & Orphans\\
\hline
N & Bias & Scatter & Bias & Scatter & Bias & Scatter & Fraction\\
 & $\langle \ln(\sigma^v/\sigma_{dm})\rangle$ & $\langle \ln(\sigma^v/\sigma_{dm})^2 \rangle^{1/2}$ & $\langle \ln(M_v/M_{200})\rangle$ & $\langle \ln(M_v/M_{200})^2 \rangle^{1/2}$ & $\langle \ln(M_c/M_{200})\rangle$ & $\langle \ln(M_c/M_{200})^2 \rangle^{1/2}$ & \\
\hline
150 & -0.02$\pm$0.02 & 0.15 & -0.03$\pm$0.05 & 0.48 & -0.05$\pm$0.04 & 0.33 & 0.55\\
100 & -0.03$\pm$0.02 & 0.16 & -0.06$\pm$0.05 & 0.51 & -0.08$\pm$0.04 & 0.36 & 0.52\\
75 & -0.04$\pm$0.02 & 0.18 & -0.09$\pm$0.05 & 0.55 & -0.11$\pm$0.04 & 0.39 & 0.51\\
50 & -0.06$\pm$0.02 & 0.21 & -0.14$\pm$0.06 & 0.63 & -0.14$\pm$0.04 & 0.43 & 0.49\\
35 & -0.09$\pm$0.02 & 0.22 & -0.24$\pm$0.05 & 0.67 & -0.21$\pm$0.04 & 0.49 & 0.48\\
25 & -0.13$\pm$0.02 & 0.26 & -0.36$\pm$0.07 & 0.78 & -0.30$\pm$0.05 & 0.55 & 0.46\\
15 & -0.21$\pm$0.03 & 0.32 & -0.58$\pm$0.09 & 0.96 & -0.48$\pm$0.07 & 0.71 & 0.43\\
10 & -0.27$\pm$0.04 & 0.41 & -0.74$\pm$0.12 & 1.21 & -0.59$\pm$0.09 & 0.93 & 0.39\\
\end{tabular}%}
\end{center}
\end{table}

\begin{table}
\begin{center}
%\scalebox{0.7}{
%\caption{}%\label{sample_bias&scatter_table}}
\begin{tabular}{c | cc | cc | cc}
\multicolumn{7}{c}{Subhalos} \\
\multicolumn{7}{c}{Velocity Dispersion and Caustic Mass bias/scatter} \\
 & \multicolumn{2}{c |}{${\sigma}^{v}_{N}$} & \multicolumn{2}{c}{Virial Mass} & \multicolumn{2}{c}{Caustic Mass}\\
\hline
N & Bias & Scatter & Bias & Scatter & Bias & Scatter\\
 & $\langle \ln(\sigma^v/\sigma_{dm})\rangle$ & $\langle \ln(\sigma^v/\sigma_{dm})^2 \rangle^{1/2}$ & $\langle \ln(M_v/M_{200})\rangle$ & $\langle \ln(M_v/M_{200})^2 \rangle^{1/2}$ & $\langle \ln(M_c/M_{200})\rangle$ & $\langle \ln(M_c/M_{200})^2 \rangle^{1/2}$\\
\hline
150 & 0.13$\pm$0.02 & 0.21 & 0.35$\pm$0.05 & 0.63 & 0.31$\pm$0.04 & 0.41\\
100 & 0.10$\pm$0.02 & 0.21 & 0.31$\pm$0.05 & 0.64 & 0.25$\pm$0.04 & 0.39\\
75 & 0.08$\pm$0.02 & 0.21 & 0.26$\pm$0.05 & 0.63 & 0.20$\pm$0.04 & 0.40\\
50 & 0.04$\pm$0.02 & 0.21 & 0.16$\pm$0.05 & 0.65 & 0.13$\pm$0.04 & 0.43\\
35 & 0.00$\pm$0.02 & 0.23 & 0.04$\pm$0.06 & 0.69 & 0.04$\pm$0.04 & 0.47\\    
25 & -0.04$\pm$0.03 & 0.27 & -0.09$\pm$0.08 & 0.81 & -0.05$\pm$0.06 & 0.54\\
15 & -0.13$\pm$0.03 & 0.33 & -0.33$\pm$0.09 & 1.01 & -0.23$\pm$0.06 & 0.69\\
10 & -0.21$\pm$0.04 & 0.41 & -0.58$\pm$0.11 & 1.23 & -0.41$\pm$0.07 & 0.83\\
\end{tabular}%}
%\label{tab:bias_scatter}
\end{center}
\end{table}

\end{document}